\documentclass[%
print,
 amsmath,amssymb,
 aps,twocolumn
]{revtex4}

\usepackage{times}
\usepackage{graphicx}
\usepackage{dcolumn}
\usepackage{bm}
\usepackage{dsfont}
\usepackage{mathrsfs}
\usepackage[landscape,papersize={297.1mm,210mm},left=1.9cm,right=1.6cm,top=2.7cm,bottom=2.8cm]{geometry}
\usepackage[                   
            pdfstartview=FitH,
            colorlinks, 
            pdfborder=001,   
            linkcolor=blue,
            anchorcolor=blue,
            citecolor=blue,
            urlcolor=blue
            ]{hyperref}
\usepackage{amsthm}
\makeatletter

\newcommand{\Rmnum}[1]{\expandafter\@slowromancap\romannumeral #1@}
\newcommand{\tabincell}[2]{\begin{tabular}{@{}#1@{}}#2\end{tabular}}
\makeatother

\begin{document}
\title{Orthogonal product sets with strong quantum nonlocality on plane structure}

\author{Huaqi Zhou$^1$}

\author{Ting Gao$^1$}
\email{gaoting@hebtu.edu.cn}

\author{Fengli Yan$^2$}
\email{flyan@hebtu.edu.cn}
\affiliation{$^1$ School of Mathematical Sciences,
 Hebei Normal University, Shijiazhuang 050024, China \\
$^2$ College of Physics, Hebei Key Laboratory of Photophysics Research and Application, Hebei Normal University, Shijiazhuang 050024, China}

\begin{abstract}
In this paper, we consider the orthogonal product set (OPS) with strong quantum nonlocality in multipartite quantum systems. Based on the decomposition of plane geometry, we present a sufficient condition for the triviality of orthogonality-preserving POVM on fixed subsystem and partially answer an open question given by Yuan et al. [\href{https://journals.aps.org/pra/abstract/10.1103/PhysRevA.102.042228} {Phys. Rev. A \textbf{102}, 042228 (2020)}]. The connection between the nonlocality and the plane structure of OPS is established. We successfully construct a strongly nonlocal OPS in $\mathcal{C}^{d_{A}}\otimes \mathcal{C}^{d_{B}}\otimes \mathcal{C}^{d_{C}}$ $(d_{A},d_{B},d_{C}\geq 4)$, which contains fewer quantum states, and generalize the structures of known OPSs to any possible three and four-partite systems. In addition, we propose several entanglement-assisted protocols for perfectly local discriminating the sets. It is shown that the protocols without teleportation use less entanglement resources on average and these sets can always be discriminated locally with multiple copies of 2-qubit maximally entangled states. These results also exhibit nontrivial signification of maximally entangled states in the local discrimination of quantum states.\\

\end{abstract}

\maketitle

\section{Introduction}
Quantum nonlocality, as one fundamental property and the most celebrated manifestations of quantum mechanics arises from entangled states. Quantum entanglement has received  extensive attention, and many results have been obtained \cite{Horodecki, Zhou11, Gao}. Since entangled pure states violate Bell type inequalities, so they are nonlocal \cite{Bell,Clauser,Freedman,Yan,Meng,Chen,Ding1,Ding2}. However, in 1999, Bennett et al. \cite{Bennett} proposed a complete orthogonal product bases with nonlocality, i.e., each of which cannot be reliably discriminated by local operations and classical communication (LOCC) while it only can be identified by a global measurement. It means that nonlocal properties are no longer restricted only to entangled systems. Later, this phenomenon, quantum nonlocality without entanglement, has aroused a wide research \cite{Niset,Zhang1,Wang1,Wang2,Feng,Xu,Zhang3,Halder1,Jiang}. Zhang et al. \cite{Zhang1} gave a class of nonlocal orthogonal product bases in the quantum system of $\mathcal{C}^{d}\otimes \mathcal{C}^{d}$, where $d$ is odd. Wang et al. \cite{Wang1} obtained a small set with only $3(m+n)-9$ orthogonal product states in an arbitrary bipartite quantum system $\mathcal{C}^{m}\otimes \mathcal{C}^{n}$ and proved that these states are LOCC indistinguishable. Xu et al. \cite{Xu} presented a locally indistinguishable set of multipartite orthogonal product states of size $2n$, which can be projected to quantum system $\otimes_{i=1}^{n}\mathcal{C}^{2}$ in essence. Jiang et al. \cite{Jiang} proposed a simple method to construct a nonlocal set of orthogonal product states in $\otimes_{i=1}^{n}\mathcal{C}^{d_{i}}(n\geq 3,d_{i}\geq 2)$ quantum system. It is also shown that local indistinguishability is a crucial primitive for quantum data hiding \cite{Terhal,DiVincenzo,Eggeling} and quantum secret sharing \cite{Hillery,Guo,Hsu,Markham,Rahaman,JWang}.

Recently, the concept of quantum nonlocality without entanglement was further developed \cite{Halder,Zhang2,Rout,Yuan,Shi1,Shi2,Shi3,Che,LiM,Bhunia,He}. Halder et al. \cite{Halder} presented a stronger manifestation of this kind of nonlocality in multiparty systems. Specifically, an orthogonal product set (OPS) on $\otimes_{i=1}^{n}\mathcal{C}^{d_{i}}(n\geq 3,d_{i}\geq 3)$ is defined to be strongly nonlocal if it is locally irreducible in every bipartition. The local irreducibility means that it is not possible to eliminate one or more states from the set by orthogonality-preserving local measurements \cite{Halder}. Immediately, Zhang et al. \cite{Zhang2} gave a more general definition of strong quantum nonlocality for multipartite quantum states, where the set is strongly nonlocal if it is locally irreducible in every $(n-1)$-partition. Naturally, the set of orthogonal quantum states which is locally irreducible in every bipartition is the strongest manifestation of nonlocality.

It is well known that entanglement is a very valuable resource which allows remote parties to communicate \cite{Gao3,Gao4} as in teleportation \cite{Bennett1,Gao1,Gao2}. In fact, the set of orthogonal quantum states with quantum nonlocality can always be perfectly discriminated by sharing additional entangled resource among the parties \cite{Rout,Cohen1,Cohen2,Bandyopadhyay,Zhang4,Li,Zhang5}. Most generally, by using enough entanglement resource, we can teleport the full multipartite states to one of the parties by LOCC, then these states can be determined by performing suitable measurement. In 2008, Cohen \cite{Cohen2} proposed protocols using entanglement more efficiently than teleportation to distinguish certain classes of unextendible product bases (UPBs), where less entanglement were consumed in comparison to the teleportation-based method. Rout et al. \cite{Rout} studied local state discrimination protocols with Einstein-Podolsky-Rosen (EPR) state and  Greenberger-Horne-Zeilinger (GHZ) state. Zhang et al. \cite{Zhang4,Zhang5} presented several protocols to locally distinguish particular UPBs by using different entanglement resource and proved that some sets can also be locally distinguished with multiple copies of EPR states.

In this paper, we investigate the OPSs with strong nonlocality. In Sec. \ref{Q2}, we introduce some notations and required preliminary concepts and results. In Sec. \ref{Q3}, we study the sufficient condition for local irreducibility of OPS and illustrate the smallest size of OPS under some specific constraints. Next, in Sec. \ref{Q4}, we generalize the structure of given sets to higher dimension systems and construct a smaller OPS with the strongest quantum nonlocality in $\mathcal{C}^{d_{A}}\otimes \mathcal{C}^{d_{B}}\otimes \mathcal{C}^{d_{C}}$ $(d_{A},d_{B},d_{C}\geq 4)$. Furthermore, we also investigate local distinguishability of our OPSs by using different entanglement resource in Sec. \ref{Q5}. Finally, we conclude  with a brief summary in Sec. \ref{Q6}.

\section{Preliminaries}\label{Q2}
In this section, we introduce some definitions and notations needed in the rest of the paper.

\emph{Definition 1} \cite{Walgate}. A measurement is trivial if all the POVM elements are proportional to the identity operator. Otherwise, the measurement is nontrivial.

In an $n$-partite system, a set $\{|\varphi\rangle\}$ of orthogonal states is locally irreducible if the orthogonality-preserving POVM \cite{Halder} on any party can only be trivial. The inverse does not hold in general. Let $X_{1}=\{2,3,\ldots,n\}$, $X_{2}=\{3,\ldots,n,1\}$, $X_{3}=\{4,\ldots,n,1,2\},\ldots,~X_{n}=\{1,2,\ldots,n-1\}$.

\emph{Lemma~1} \cite{Shi2}. If $X_{i}$ party can only perform a trivial orthogonality-preserving POVM for all $1\leq i\leq n$, then the set $\{|\varphi\rangle\}$ is of the strongest nonlocality \cite{Zhang2}.

Let the $d\times d$ matrix $E=(a_{ij})_{i,j\in\mathcal{Z}_{d}}$ be the matrix representation of the operator $E=M^{\dagger}M$ in the basis $\mathcal{B}=\{|0\rangle,\ldots,|d-1\rangle\}$.  Define
\begin{equation}
_{\mathcal{S}}E_{\mathcal{T}}=\sum_{|i\rangle\in\mathcal{S}}\sum_{|j\rangle\in\mathcal{T}}a_{ij}|i\rangle\langle j|,
\end{equation}
where $\mathcal{S}$ and $\mathcal{T}$ are two nonempty subsets of $\mathcal{B}$.
Especially,  $_{\mathcal{T}}E_{\mathcal{T}}$ is represented by $E_{\mathcal{T}}$. Let $\{|\psi_{i}\rangle\}_{i=0}^{s-1}$ and $\{|\phi_{j}\rangle\}_{j=0}^{t-1}$ be two orthogonal sets spanned by $\mathcal{S}$ and $\mathcal{T}$, respectively, where $s=|\mathcal{S}|$ and $t=|\mathcal{T}|$.

\emph{Lemma~2} \cite{Shi2}. If subsets $\mathcal{S}$ and $\mathcal{T}$ are disjoint and $\langle\psi_{i}|E|\phi_{j}\rangle=0$ for any $i\in\mathcal{Z}_{s}$, $j\in\mathcal{Z}_{t}$, then $_{\mathcal{S}}E_{\mathcal{T}}=\textbf{0}$ and $_{\mathcal{T}}E_{\mathcal{S}}=\textbf{0}$.

\emph{Lemma~3} \cite{Shi2}. Suppose that $\langle\psi_{i}|E|\psi_{j}\rangle=0$ for any $i\neq j\in\mathcal{Z}_{s}$. If there exists a state $|i_{0}\rangle\in \mathcal{S}$ such that $_{\{|i_{0}\rangle\}}E_{\mathcal{S}\setminus\{|i_{0}\rangle\}}=\textbf{0}$ and $\langle i_{0}|\psi_{j}\rangle\neq 0$ for any $j\in\mathcal{Z}_{s}$, then $E_{\mathcal{S}}\propto \mathcal{I}_{s}$, i.e. $E_{\mathcal{S}}$ is proportional to the identity matrix.

Consider an $n$-partite quantum system $\mathcal{H}=\otimes_{i=1}^{n}\mathcal{C}^{d_{i}}$. The computational basis of the whole quantum system is denoted by $\mathcal{B}=\{|i\rangle\}_{i=0}^{d_1d_2\cdots d_n -1}=\{\otimes_{k=1}^{n}|i_{k}\rangle ~ | ~ i_k=0, 1, \cdots, d_{k}-1 \}=\mathcal{B}^{\{1\}}\otimes \mathcal{B}^{\{2\}}\otimes \cdots\otimes \mathcal{B}^{\{n\}}$, where $\mathcal{B}^{\{k\}}=\{|i_{k}\rangle\}_{{i_k}=0}^{d_{k}-1}$ is the computational basis of the $k$th subsystem. Let
\begin{equation}
\mathcal{B}_{r}=\mathcal{B}_{r}^{\{1\}}\otimes \mathcal{B}_{r}^{\{2\}}\otimes \cdots\otimes \mathcal{B}_{r}^{\{n\}}
\end{equation}
be a subset of basis $\mathcal{B}$ with $\mathcal{B}_{r}^{\{i\}}\subset \mathcal{B}^{\{i\}}$. Suppose that $\mathcal{B}_{r}$ ($1\leq r \leq q$) are  disjoint subsets of $\mathcal{B}$,
then, there is a class of OPS,
\begin{equation}\label{5}
\begin{aligned}
S=\cup_{r\in Q}S_{r},~~Q=\{1,2, \ldots,q\},
\end{aligned}
\end{equation}
in $\mathcal{H}$, where $S_{r}$ expresses the orthogonal product basis of the subspace  spanned by  $\mathcal{B}_{r}$, and each component of the vector in $S_{r}$ is nonzero under the computational basis  $\mathcal{B}_r$, that is,
each vector $|\phi\rangle_{r}$ in $S_{r}$ has the following form
\begin{equation}\label{phi-r}
\begin{aligned}
|\phi\rangle_{r}=\bigg(\sum_{|j_{1}\rangle\in \mathcal{B}_{r}^{\{1\}}}a_{j_{1}}^{(1)}|j_{1}\rangle\bigg)\otimes\cdots\otimes\bigg(\sum_{|j_{n}\rangle\in \mathcal{B}_{r}^{\{n\}}}a_{j_{n}}^{(n)}|j_{n}\rangle\bigg)
\end{aligned}
\end{equation}
with nonzero complex numbers $a_{j_{k}}^{(k)}$ for $k=1, 2, \ldots,n$. If the set $S$ is invariant under cyclic permutation of all subsystems, then we call it symmetric.

A plane structure of the set $S$ refers to a two-dimensional grid diagram and each subset $S_{r}$ corresponds a domain in the diagram.

\begin{figure}[h]
\centering
\includegraphics[width=0.24\textwidth]{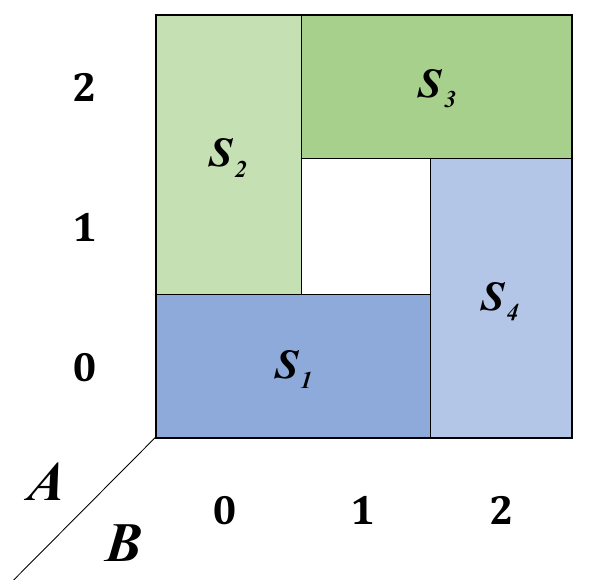}
\caption{The plane structure of OPS given by Eq. (\ref{2}) in bipartition. \label{8}}
\end{figure}

\emph{Example 1}. In $\mathcal{C}^{3}\otimes\mathcal{C}^{3}$, let
\begin{equation}\label{2}
\begin{aligned}
&S_{1}=|0\rangle|0\pm 1\rangle, && S_{2}=|1\pm 2\rangle|0\rangle,\\
&S_{3}=|2\rangle|1\pm 2\rangle, && S_{4}=|0\pm 1\rangle|2\rangle,
\end{aligned}
\end{equation}
the plane structure of the OPS \cite{Bennett} $S=\cup_{i=1}^4 S_i$ is depicted in Fig.\ref{8}. The four dominos in this geometry structure represent the four subsets $S_{1}$, $S_{2}$, $S_{3}$ and $S_{4}$, respectively.

In order to facilitate the establishment of the connection between the nonlocality and the plane structure of the given set $S$, some symbols are introduced. Given a subset $X$ of $\{1,2,\ldots,n\}$ and its complement $Y=\bar{X}$, we use $\mathcal{B}^{X}=\{|i\rangle_{X}\}_{i=0}^{d_{X}-1}$ with $d_X=\prod_{j\in X} d_j$ to represent the computation basis of the Hilbert space $\mathcal{H}_X=\otimes_{j\in X} \mathcal{C}^{d_j}$ corresponding to $X$ party and analogously  $\mathcal{B}^{Y}=\{|i\rangle_{Y}\}_{i=0}^{d_{Y}-1}$ corresponding to $Y$ party. Under the basis $\mathcal{B}$, the projection set of $S_{r}$ on $\tau$ $(\tau=X,Y)$ party is expressed as
$S_{r}^{(\tau)}=\{\textup{Tr}_{\bar{\tau}}(|i\rangle\langle i|)~|~|i\rangle\in \mathcal{B}~\textup{and}~\langle i|\phi^{r}\rangle\neq 0~\textup{for~any}~|\phi^{r}\rangle\in S_{r}\}$. Naturally, the projection set $S_{r}^{(\tau)}$ is a subset of basis $\mathcal{B}^{\tau}$. For a fixed $i\in\mathcal{Z}_{d_{X}}$, let $\mathcal{B}_{i}^{X}:=\{|k\rangle_{X}\}_{k=i}^{d_{X}-1}$, $V_{i}:=\{\bigcup_{v}S_{v}^{(Y)}~|~|i\rangle_{X}\in S_{v}^{(X)}\}$ and $\widetilde{S}_{V_{i}}:=\{\bigcup_{j}S_{j}^{(X)}~|~S_{j}^{(Y)}\cap V_{i}\neq \emptyset\}$.

\emph{Example 2}. Consider the OPS given by Eq. (\ref{2}). $X$ and $Y$ represent $B$ and $A$, respectively. Observe its plane structure as shown in Fig. \ref{8}, the projection set of a subset on the $B$ $(\textup{or}~A)$ party is actually the coordinate of the corresponding grid on the $B$ $(\textup{or}~A)$ party. We have
\begin{equation}\label{3}
\begin{aligned}
&S_{1}^{(B)}=\{|0\rangle_{B},|1\rangle_{B}\}, && S_{2}^{(B)}=\{|0\rangle_{B}\},\\
&S_{3}^{(B)}=\{|1\rangle_{B},|2\rangle_{B}\}, && S_{4}^{(B)}=\{|2\rangle_{B}\},\\
\end{aligned}
\end{equation}
and
\begin{equation}\label{4}
\begin{aligned}
&S_{1}^{(A)}=\{|0\rangle_{A}\}, && S_{2}^{(A)}=\{|1\rangle_{A},|2\rangle_{A}\},\\
&S_{3}^{(A)}=\{|2\rangle_{A}\}, && S_{4}^{(A)}=\{|0\rangle_{A},|1\rangle_{A}\}.\\
\end{aligned}
\end{equation}

For all $i\in\mathcal{Z}_{3}$, $\mathcal{B}_{i}^{B}$ is a subset of basis $\mathcal{B}^B$ and $\mathcal{B}_{0}^{B}$ is equal to $\mathcal{B}^B$. It is easy to know that
\begin{equation*}
\begin{aligned}
&\mathcal{B}_{0}^{B}=\{|0\rangle_{B},|1\rangle_{B},|2\rangle_{B}\},\\
&\mathcal{B}_{1}^{B}=\{|1\rangle_{B},|2\rangle_{B}\},\\
&\mathcal{B}_{2}^{B}=\{|2\rangle_{B}\}.
\end{aligned}
\end{equation*}

Since $V_{i}$ expresses the union of the projection sets $S_{v}^{(A)}$ of  $S_v$ on $A$ party, where all corresponding  projection sets $S_{v}^{(B)}$ of $S_v$ on $B$ party contain quantum state $|i\rangle_{B}$, then there are
\begin{equation*}
\begin{aligned}
&V_{0}=S_{1}^{(A)}\cup S_{2}^{(A)}=\{|0\rangle_{A},|1\rangle_{A},|2\rangle_{A}\},\\
&V_{1}=S_{1}^{(A)}\cup S_{3}^{(A)}=\{|0\rangle_{A},|2\rangle_{A}\},\\
&V_{2}=S_{3}^{(A)}\cup S_{4}^{(A)}=\{|0\rangle_{A},|1\rangle_{A},|2\rangle_{A}\}.
\end{aligned}
\end{equation*}
Note that each projection set $S_j^{(A)}$ contains a quantum state in $V_{i}$, $\widetilde{S}_{V_{i}}$ is the union of all the projection sets $S_j^{(B)}$ of $S_j$ on $B$ party. That is,
\begin{equation*}
\begin{aligned}
&\widetilde{S}_{V_{0}}=\widetilde{S}_{V_{1}}=\widetilde{S}_{V_{2}}=\cup_{j=1}^{4}S_{j}^{(B)}=\{|0\rangle_{B},|1\rangle_{B},|2\rangle_{B}\}.
\end{aligned}
\end{equation*}

\emph{Definition 2}. A family of projection sets $\{S_{r}^{(\tau)}\}_{r\in Q}$ is connected if it cannot be divided into two groups of sets $\{S_{k}^{(\tau)}\}_{k\in T}$ $(T\subsetneqq Q)$ and $\{S_{l}^{(\tau)}\}_{l\in Q\setminus T}$ such that
\begin{equation}\label{1}
\begin{aligned}
\bigg(\bigcup\limits_{k\in T}S_{k}^{(\tau)}\bigg)\bigcap \bigg(\bigcup\limits_{l\in Q\setminus T}S_{l}^{(\tau)}\bigg)=\emptyset.
\end{aligned}
\end{equation}

\emph{Definition 3}. $R_{r}=\bigcup_{k\in T}S_{k}$ $(r\notin T\subset Q)$ is called the projection inclusion (PI) set of $S_{r}$ on $X$ party if projection sets satisfy $\bigcap_{k\in T}S_{k}^{(Y)}\neq \emptyset$ and $S_{r}^{(X)}\subset \bigcup_{k\in T} S_{k}^{(X)}$.
Specially, $R_{r}$ is called a more useful projection inclusion (UPI) set if there exists a subset $S_{k}\subset R_{r}$ such that $|S_{r}^{(X)}\bigcap S_{k}^{(X)}|=1$.

From the definition, both the PI set and the UPI set of a subset $S_{r}$ of an OPS $S$ may not be unique.
By observing the plane tile as shown in Fig. \ref{8}, it is easy to know that both $S_1$ and $S_1\cup S_4$ are PI sets of $S_2$ in (\ref{2}) on $B$ party, and
 $S_{2}\cup S_{3}$ is the PI set of $S_{1}$ on $B$ party. Due to $|S_{1}^{(B)}\cap S_{2}^{(B)}|=1$, these PI sets are also UPI sets.

For the set $S$ in (\ref{5}), we construct a set sequence $G_{1},G_{2},\ldots,G_{s}$. The set $G_{1}$ denoted as $\cup_{r_{1}\in T_{1}}S_{r_{1}}$ is the union of all subsets $S_{r_{1}}$  that have UPI sets. The remaining sets $G_{2},\ldots,G_{s}$ are expressed by $\cup_{r_{2}\in T_{2}}S_{r_{2}},\ldots,\cup_{r_{s}\in T_{s}}S_{r_{s}}$, respectively. Moreover, this sequence also satisfies the following two conditions.

1) The sets $G_{x}$ are pairwise disjoint and the union of all sets is $S$.

2) For any $S_{r_{x+1}}\subset G_{x+1}$ $(x=1,\ldots,s-1)$, there is always a subset $S_{r_{x}}\subset G_{x}$ such that $S_{r_{x}}^{(X)}\cap S_{r_{x+1}}^{(X)}\neq\emptyset$.

Note that such a set sequence $G_{1},G_{2},\ldots,G_{s}$ satisfying above 1) and 2) does not necessarily exist.
In addition, we call $S_{r_{x}}$ an included (IC) subset about set $G_{x}$ $(x=1,\ldots,s)$, if there is a subset $S_{r_{x}'}\subset G_{x}$ such that $S_{r_{x}}^{(X)}\subsetneqq S_{r_{x}'}^{(X)}$. Otherwise, it is called a non-included (NIC) subset.

\emph{Example 3}. We consider the OPS in (\ref{2}), where each subset has a corresponding UPI set
\begin{equation}
\begin{aligned}
&R_{1}=S_{2}\cup S_{3},~ R_{2}=S_{1},~ R_{3}=S_{1}\cup S_{4},~ R_{4}=S_{3}.
\end{aligned}
\end{equation}
So, there is only one set in its set sequence, which happens to be this OPS. That is $G_{1}=\cup_{i=1}^{4}S_{i}$.

\section{The sufficient condition for the triviality of orthogonality-preserving
POVM and the smallest size of OPS under some constraints}\label{Q3}

It is an important way to illustrate the irreducibility of OPS by proving that the orthogonality-preserving POVM on the subsystems can only be trivial \cite{Jiang,Halder,Zhang2,Shi2,Yuan,Bhunia}. Here, we will present a sufficient condition for orthogonality-preserving POVM being triviality. On plane structure, the condition is efficient for constructing OPS with strong nonlocality and demonstrating the irreducibility of given OPS.

\emph{Theorem~1}. For the given set $S$ in (\ref{5}), any orthogonality-preserving POVM performed on $X$ party can only be trivial if the following conditions are satisfied.

i) There is an inclusion relationship $\mathcal{B}_{i}^{X}\subset \widetilde{S}_{V_{i}}$ for any $i\in\mathcal{Z}_{d_{X}-1}$.

ii) For any subset $S_{r}$, there exists a corresponding PI set $R_{r}$ on $X$ party.

iii) There is a set sequence $G_{1},\ldots,G_{s}$ satisfying 1) and 2). Moreover, for each NIC subset $S_{r_{x+1}}\subset G_{x+1}$, there exist a subset $S_{r_{x}}\subset G_{x}$ and a subset $S_{r_{x+1}'}\subset R_{r_{x+1}}$ such that $S_{r_{x}}^{(X)}\cap S_{r_{x+1}}^{(X)}\supset S_{r_{x+1}}^{(X)}\cap S_{r_{x+1}'}^{(X)}$ with $x=1,2,\ldots,s-1$.

iv) The family of sets $\{S_{r}^{(X)}\}_{r\in Q}$ is connected.

$Proof.$ Let $\{E\}$ be an any orthogonality-preserving POVM performed on $X$. Without loss of generality, we assume
\begin{equation}
\begin{aligned}
E=\begin{pmatrix} a_{00} & a_{01} & \cdots & a_{0(d_{X}-1)} \\
                      a_{10} & a_{11} & \cdots & a_{1(d_{X}-1)} \\
                      \vdots & \vdots & \ddots & \vdots \\
                      a_{(d_{X}-1)0} & a_{(d_{X}-1)1} & \cdots & a_{(d_{X}-1)(d_{X}-1)} \end{pmatrix},
\end{aligned}
\end{equation}
in the computation basis $\mathcal{B}^{X}$. Because the postmeasurement states should be mutually orthogonal, for any two states $|\psi_{1}\rangle_{X}|\phi_{1}\rangle_{Y}$ and $|\psi_{2}\rangle_{X}|\phi_{2}\rangle_{Y}$ in $S$, we have $_{X}\langle\psi_{1}|_{Y}\langle\phi_{1}|E\otimes \mathcal{I}|\psi_{2}\rangle_{X}|\phi_{2}\rangle_{Y}=0$. If $\langle\phi_{1}|\phi_{2}\rangle_{Y}\neq 0$, then $_{X}\langle\psi_{1}|E|\psi_{2}\rangle_{X}=0$.

\begin{figure}[h]
\begin{minipage}[t]{0.4\linewidth}
\centering
\includegraphics[width=1\textwidth]{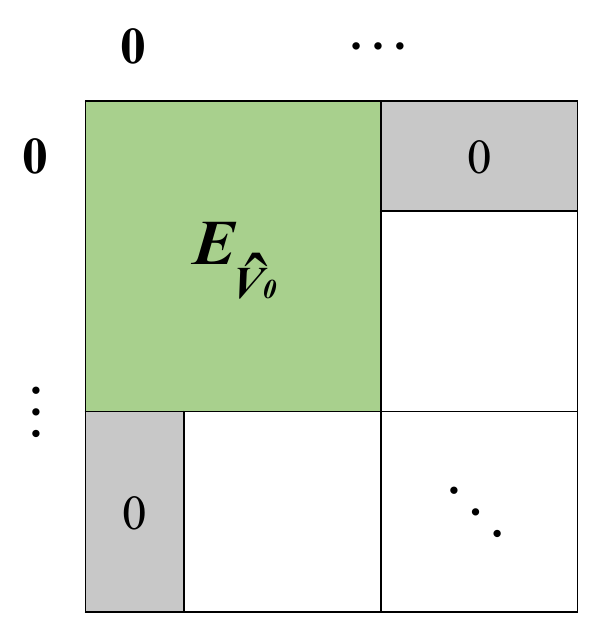}
\caption{In step 1, taking $i=0$ as an example, we show that all the elements of $E$ in the first row and in the first column except $E_{\widehat{V}_{0}}$ are zero. \label{51}}
\end{minipage}
\hfill
\begin{minipage}[t]{0.45\linewidth}
\centering
\includegraphics[width=1\textwidth]{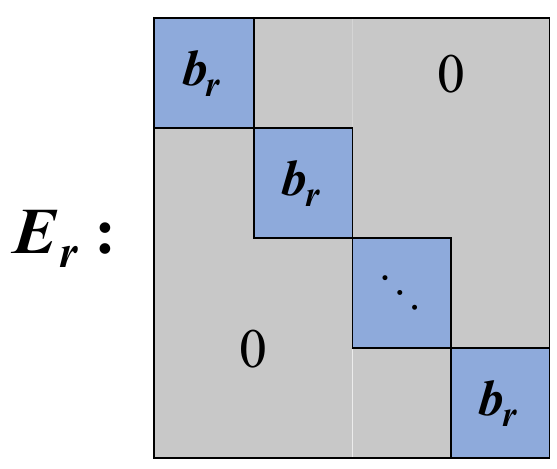}
\caption{In steps 2 and 3, it is proved that the operator $E_{r}=E_{S_r^{(X)}}$ corresponding to subset $S_{r}$ is proportional to the unit operator for all $r\in Q$. \label{52}}
\end{minipage}\\
\begin{minipage}[t]{0.4\linewidth}
\centering
\includegraphics[width=1\textwidth]{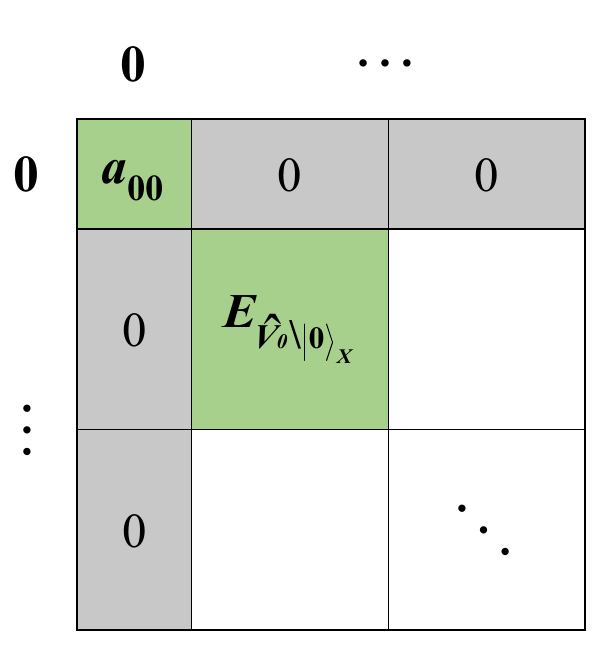}
\caption{Consider the operator $E_{\widehat{V}_{0}}$. Because each $E_{v}\propto \mathcal{I}$, only element $a_{00}$ in the first row is nonzero. We can get  the similar result for other $\widehat{V}_{i}$ $(i=1,\ldots,d_{X}-2)$. Therefore, we deduce that the off-diagonal elements of $E$ are all zero. \label{53}}
\end{minipage}
\hfill
\begin{minipage}[t]{0.45\linewidth}
\centering
\includegraphics[width=1\textwidth]{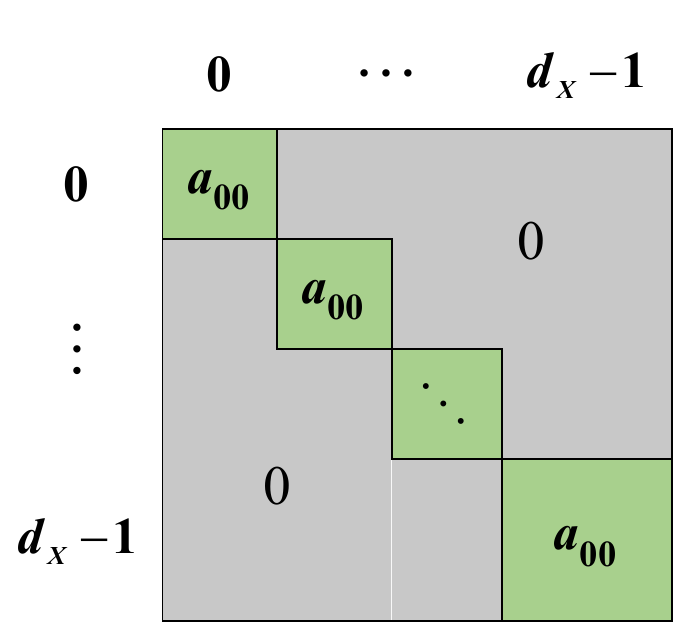}
\caption{It follows from  condition iv) that the scalars $b_r$ are all equal. Then the diagonal entries of the POVM element $E$ are all equal, that is, $E=a_{00}\mathcal{I}$ for some positive real number $a_{00}$, where $\mathcal{I}$ is the identity matrix. \label{54}}
\end{minipage}
\end{figure}

Let $S_{r}^{\tau}=\{\textup{Tr}_{\bar{\tau}}(|\phi^{r}\rangle\langle\phi^{r} |)~|~|\phi^{r}\rangle\in S_{r}\}$ $(\tau=X,Y)$ express the set of reduced density matrices. For any two different subsets $S_{q_{1}}$ and $S_{q_{2}}$, if $S_{q_{1}}^{(Y)}\cap S_{q_{2}}^{(Y)}\neq \emptyset$, then $S_{q_{1}}^{(X)}\cap S_{q_{2}}^{(X)}= \emptyset$ and there always exist two states $|\phi_{q_{1}}\rangle_{Y}\in S_{q_{1}}^{Y}$ and $|\phi_{q_{2}}\rangle_{Y}\in S_{q_{2}}^{Y}$ such that $\langle \phi_{q_{1}}|\phi_{q_{2}}\rangle_{Y}\neq 0$. Due to the orthogonality-preserving property, we obtain $_{X}\langle \psi_{q_{1}}|E|\psi_{q_{2}}\rangle_{X}=0$ for all $|\psi_{q_{1}}\rangle_{X}\in S_{q_{1}}^{X}$ and $|\psi_{q_{2}}\rangle_{X}\in S_{q_{2}}^{X}$. According to Lemma 2, we deduce  $_{S_{q_{1}}^{(X)}}E_{S_{q_{2}}^{(X)}}=\textbf{0}$. Using this result, we can prove that $E\propto \mathcal{I}$ by the following four steps. Here, figures \ref{51}-\ref{54} depict the process of proving.

\emph{Step 1}. When $i=0$, we know $V_{0}=\{\cup_{v}S_{v}^{(Y)}~|~|0\rangle_{X}\in S_{v}^{(X)}\}$. For each $|j\rangle_{Y}\in V_{0}$, let $\{S_{j_{s}}\}_{j_{s}\in Q_{j}}$ $(Q_{j}\subset Q)$ represent the all subsets whose projection sets on $Y$ party contain the state $|j\rangle_{Y}$. Suppose $S_{j_{1}}$ is the subset such that $|0\rangle_{X}\in S_{j_{1}}^{(X)}$, then one has $_{S_{j_{1}}^{(X)}}E_{S_{j_{s}}^{(X)}}=\textbf{0}$ for any $s$ $(s\neq 1)$. By the definition and condition i), it is easy to derive  $\cup_{j,s}S_{j_{s}}^{(X)}=\widetilde{S}_{V_{0}}=\mathcal{B}_{0}^{X}=\{|i\rangle_{X}\}_{i=0}^{d_{X}-1}$. Thus, we get $a_{0k_{0}}=a_{k_{0}0}=0$ for $|k_{0}\rangle_{X}\notin \widehat{V}_{0}$, where $\widehat{V}_{0}=\{\cup_{v}S_{v}^{(X)}~|~|0\rangle_{X}\in S_{v}^{(X)}\}$. See Fig. \ref{51}.

Similarly, when $i=1,\ldots,d_{X}-2$, we obtain $a_{ik_{i}}=a_{k_{i}i}=0$ for $|k_{i}\rangle_{X}\notin \widehat{V}_{i}$ and $k_{i}>i$. Here $\widehat{V}_{i}=\{\cup_{v}S_{v}^{(X)}~|~|i\rangle_{X}\in S_{v}^{(X)}\}$.

\emph{Step 2}. According to the condition ii), for each $r\in Q$, there exists a PI set $R_{r}=\bigcup_{t\in T_r}S_{t}$ ($r\notin T_r\subset Q$) of $S_{r}$ on $X$ party, where $\bigcap_{t\in T_r}S_{t}^{(Y)}\neq \emptyset$ and $S_{r}^{(X)}\subset \bigcup_{t\in T_r} S_{t}^{(X)}$. For any two different indexes $t_1$ and $t_2$ in $T_r$, it is not difficult to deduce that $a_{kl}=a_{lk}=0$ with $|k\rangle_{X}\in S_{t_1}^{(X)}\cap S_{r}^{(X)}$ and $|l\rangle_{X}\in S_{t_2}^{(X)}\cap S_{r}^{(X)}$ for $k\neq l$.

\emph{Step 3}. For any subset $S_{r_{1}}$ in $G_{1}$, the corresponding set $R_{r_{1}}$ is  UPI set. From Definition 3, there is a subset $S_{r_{1}'}\subset R_{r_{1}}$ such that $|S_{r_{1}}^{(X)}\cap S_{r_{1}'}^{(X)}|=1$. It is a special case in the step 2. Let $|k\rangle_{X}$ be the only one element of $S_{r_{1}}^{(X)}\cap S_{r_{1}'}^{(X)}$, then  $a_{kl}=0$ for all $|l\rangle_{X}\in S_{r_{1}}^{(X)}\setminus \{|k\rangle_{X}\}$. Since each component of the vector in $S_{r_{1}}$ is nonzero under the computation basis $\mathcal{B}_{r_1}$ from (\ref{phi-r}), it is easy to know $\langle k|\psi\rangle_{X}\neq 0$ for any $|\psi\rangle_{X}\in S_{r_{1}}^{X}$. According to Lemma 3, we deduce $E_{r_{1}}=E_{S_{r_{1}}^{(X)}}\propto \mathcal{I}$.

By the condition iii), for each NIC subset $S_{r_{2}}\subset G_{2}$, there exist a subset $S_{r_{1}}\subset G_{1}$ and a subset $S_{r_{2}'}\subset R_{r_{2}}$ such that $S_{r_{1}}^{(X)}\cap S_{r_{2}}^{(X)}\supset S_{r_{2}}^{(X)}\cap S_{r_{2}'}^{(X)}$. Then $a_{kl}=0$ for $|k\rangle_{X},|l\rangle_{X}\in S_{r_{2}}^{(X)}\cap S_{r_{2}'}^{(X)}$ and $k\neq l$. Combining this with the step 2 produces $a_{kl}=0$ for $|k\rangle_{X}\in S_{r_{2}}^{(X)}\cap S_{r_{2}'}^{(X)}$, $|l\rangle_{X}\in S_{r_{2}}^{(X)}$ and $k\neq l$. It follows from Lemma 3 that $E_{r_{2}}=E_{S_{r_{2}}^{(X)}}\propto \mathcal{I}$. For each IC subset $S_{r_{2}''}$, there is always a corresponding NIC subset $S_{r_{2}}$ that satisfies the inclusion relationship $S_{r_{2}''}^{(X)}\subsetneqq S_{r_{2}}^{(X)}$, which implies $E_{r_{2}''}=E_{S_{r_{2}''}^{(X)}}\propto \mathcal{I}$. Similarly, $E_{r}\propto \mathcal{I}$ for each $r$. That is, there is a positive real number $b_{r}$ such that $E_{r}=b_{r}\mathcal{I}$. See also Fig. \ref{52}.

\emph{Step 4}. Consider the set $\widehat{V}_{0}=\{\cup_{v}S_{v}^{(X)}~|~|0\rangle_{X}\in S_{v}^{(X)}\}$ of step 1. Due to each $E_{v}\propto \mathcal{I}$, we have $a_{0k_{0}}=0$ for $|k_{0}\rangle_{X}\in \widehat{V}_{0}$ and $k_{0}\neq 0$. Combining this with the step 1 produces $a_{0k_{0}}=0$ for all $k_{0}>0$. We can obtain  the similar result for other $\widehat{V}_{i}$ $(i=1,\ldots,d_{X}-2)$. So, we deduce that the off-diagonal elements of $E$ are all zero. It is shown in Fig. \ref{53}. In addition, for any $x,y\in Q$, if $S_{x}^{(X)}\cap S_{y}^{(X)}\neq \emptyset$, then $b_{x}=b_{y}$. The condition iv) indicates that the family of sets $\{S_{r}^{(X)}\}_{r\in Q}$ is connected. This means that these scalars $b_{r}$ are all equal. Therefore, the POVM element can only be proportional to the unit operator $\mathcal{I}$. See also Fig. \ref{54}.  ~ \hfill $\square$

\emph{Corollary~1}. If the conditions i)-iv) in Theorem 1 are satisfied for $X=X_1, X_2, \cdots, X_n$ with $X_j=\{1, 2, \cdots, j-1, j+1, \cdots, n\}$, then  the set (\ref{5}) is an OPS of the strongest quantum  nonlocality.

Note that it is obvious that $E_{r}\propto \mathcal{I}$ for each $r\in Q$, if the set $G_{1}$ is equal to the set $S$. That is, when the set sequence  has only one set $G_{1}$, we still say that the condition iii) is valid. Next we will provide an example to show the application of this theorem on plane structure.

\emph{Example 4}. We revisit the quantum nonlocality of the following OPS \cite{Yuan} in $\mathcal{C}^{3}\otimes\mathcal{C}^{3}\otimes\mathcal{C}^{3}$
\begin{equation}\label{20}
\begin{aligned}
&S_{1}=\{|0\rangle|1\rangle|0\pm 1\rangle\}, && S_{7}=\{|0\rangle|2\rangle|0\pm 2\rangle\}, \\
&S_{2}=\{|1\rangle|0\pm 1\rangle|0\rangle\}, && S_{8}=\{|2\rangle|0\pm 2\rangle|0\rangle\}, \\
&S_{3}=\{|0\pm 1\rangle|0\rangle|1\rangle\}, && S_{9}=\{|0\pm 2\rangle|0\rangle|2\rangle\}, \\
&S_{4}=\{|1\rangle|2\rangle|0\pm 1\rangle\}, && S_{10}=\{|2\rangle|1\rangle|0\pm 2\rangle\}, \\
&S_{5}=\{|2\rangle|0\pm 1\rangle|1\rangle\}, && S_{11}=\{|1\rangle|0\pm 2\rangle|2\rangle\}, \\
&S_{6}=\{|0\pm 1\rangle|1\rangle|2\rangle\}, && S_{12}=\{|0\pm 2\rangle|2\rangle|1\rangle\}.
\end{aligned}
\end{equation}

\begin{figure}[h]
\centering
\includegraphics[width=0.46\textwidth]{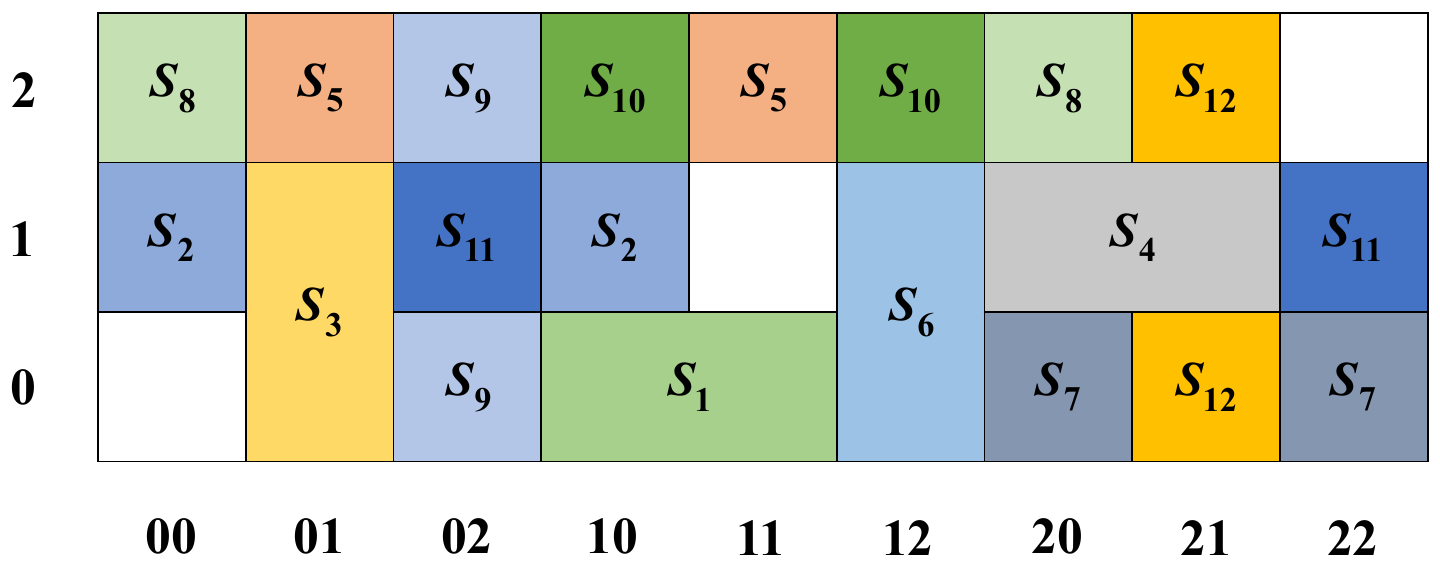}
\caption{The corresponding $3\times 9$ grid of $\{S_{r}\}_{r=1}^{12}$ given by Eq. (\ref{20}) in $A|BC$ bipartition. \label{9}}
\end{figure}

Due to Lemma 1 and the symmetry of the OPS given by Eq. (\ref{20}), we only need to consider the orthogonality-preserving POVM performed on $BC$ party. Fig. \ref{9} is the plane structure of OPS in $A|BC$ bipartition. By observing this tile graph, we can easily obtain the four conditions in Theorem 1.

First, because the projection set $\cup_{r}S_{r}^{(ABC)}$ differs from the computation basis $\mathcal{B}$ only by states $|000\rangle$, $|111\rangle$ and $|222\rangle$. It is obvious that $\widetilde{S}_{V_{ij}}=\mathcal{B}^{BC}$ for $i,j\in\mathcal{Z}_{3}$. Here $\mathcal{B}^{BC}$ is the computation basis on subsystem $BC$. Naturally, $\mathcal{B}_{ij}^{BC}\subset\widetilde{S}_{V_{ij}}$. The condition i) holds.

Second, for each subset $S_{r}$, we have the corresponding PI sets $R_{1}=S_{5}\cup S_{10}$, $R_{2}=S_{8}\cup S_{10}$, $R_{3}=S_{5}$, $R_{4}=S_{8}\cup S_{12}$, $R_{5}=S_{1}\cup S_{3}$, $R_{6}=S_{10}$, $R_{7}=S_{4}\cup S_{11}$, $R_{8}=S_{2}\cup S_{4}$, $R_{9}=S_{11}$, $R_{10}=S_{2}\cup S_{6}$, $R_{11}=S_{7}\cup S_{9}$ and $R_{12}=S_{4}$. The condition ii) is demonstrated.

Furthermore, for any two subsets $S_{x}$ and $S_{y}$, we have $|S_{x}^{(BC)}\cap S_{y}^{(BC)}|\leq 1$. So, each $R_{r}$ is an UPI set, i.e., $G_{1}$ is the union of all subsets. It is obvious that the condition iii) holds.

Finally, we find a sequence of projection sets $(S_{5}^{(BC)},~S_{10}^{(BC)})\rightarrow S_{1}^{(BC)}\rightarrow S_{2}^{(BC)}\rightarrow S_{8}^{(BC)}\rightarrow S_{4}^{(BC)}\rightarrow S_{7}^{(BC)}\rightarrow S_{11}^{(BC)}$. In this sequence, the intersection of the sets on both sides of the arrow is not empty and the union of these sets is the computation basis $\mathcal{B}^{BC}$. So, it is impossible to divide all projection sets into disjoint two groups of projection sets. That is, the family of projection sets $\{S_{r}^{(BC)}\}_{r=1}^{12}$ is connected. The condition iv) is satisfied.

According to Theorem 1, we deduce the POVM performed on $BC$ party can only be trivial. Therefore, the OPS given by Eq. (\ref{20}) is of the strongest quantum nonlocality.

For the same set as stated in Theorem 1, we have the following corollary.

\emph{Corollary~2}. If any orthogonality-preserving POVM element performed on $X$ party can only be proportional to the identity operator, then the set $\cup_{r\in Q}S_{r}^{(X)}$ is the basis $\mathcal{B}^{X}$ and the family of projection sets $\{S_{r}^{(X)}\}_{r\in Q}$ is connected.

By using Corollary 2, in systems $\mathcal{C}^{3}\otimes\mathcal{C}^{3}\otimes\mathcal{C}^{3}$ and $\mathcal{C}^{4}\otimes\mathcal{C}^{4}\otimes\mathcal{C}^{4}$, we can discuss the minimum size of the OPS given by Eq. (\ref{5}) under the specific restrictions. Let $N$ express the maximum size of all subsets, i.e., $N=\max_{r}|S_{r}|$. We have the following two theorems.

\emph{Theorem~2}. In $\mathcal{C}^{3}\otimes\mathcal{C}^{3}\otimes\mathcal{C}^{3}$, for the set $S$ (\ref{5}), if the set $S$  is symmetric and any orthogonality-preserving POVM performed on $BC$ party can only be trivial, then the set $S$ is an OPS of the strongest nonlocality.  The smallest size of this set is 24.

\emph{Theorem~3}. In $\mathcal{C}^{4}\otimes\mathcal{C}^{4}\otimes\mathcal{C}^{4}$, for the set $S$ in (\ref{5}), if $S$ is symmetric with $N=2$ and any orthogonality-preserving POVM element performed on $BC$ party can only be proportional to identity, then the set $S$ is an OPS of the strongest nonlocality.  The smallest size of this set $S$ is 48.

The detailed proofs are given in Appendix \ref{B} and \ref{C}, respectively. Theorems 2 and 3 show the minimum sizes of two kinds of OPSs with strong nolocality, respectively. They are  partial answers to an open question in Ref. \cite{Yuan}, ``Can we find the smallest strongly nonlocal set in $\mathcal{C}^{3}\otimes\mathcal{C}^{3}\otimes\mathcal{C}^{3}$, and more generally in any tripartite systems?''.

\section{OPS with the strongest quantum nonlocolity in $\mathcal{C}^{d_{A}}\otimes\mathcal{C}^{d_{B}}\otimes\mathcal{C}^{d_{C}}$ and $\mathcal{C}^{d_{A}}\otimes\mathcal{C}^{d_{B}}\otimes\mathcal{C}^{d_{C}}\otimes\mathcal{C}^{d_{D}}$}\label{Q4}

From Theorem 1, we know that the nonlocality of OPS is closely related to its plane structure. In this section, we will provide several strongly nonlocal OPSs in three and four-partite systems.

\begin{figure}[h]
\centering
\includegraphics[width=0.46\textwidth]{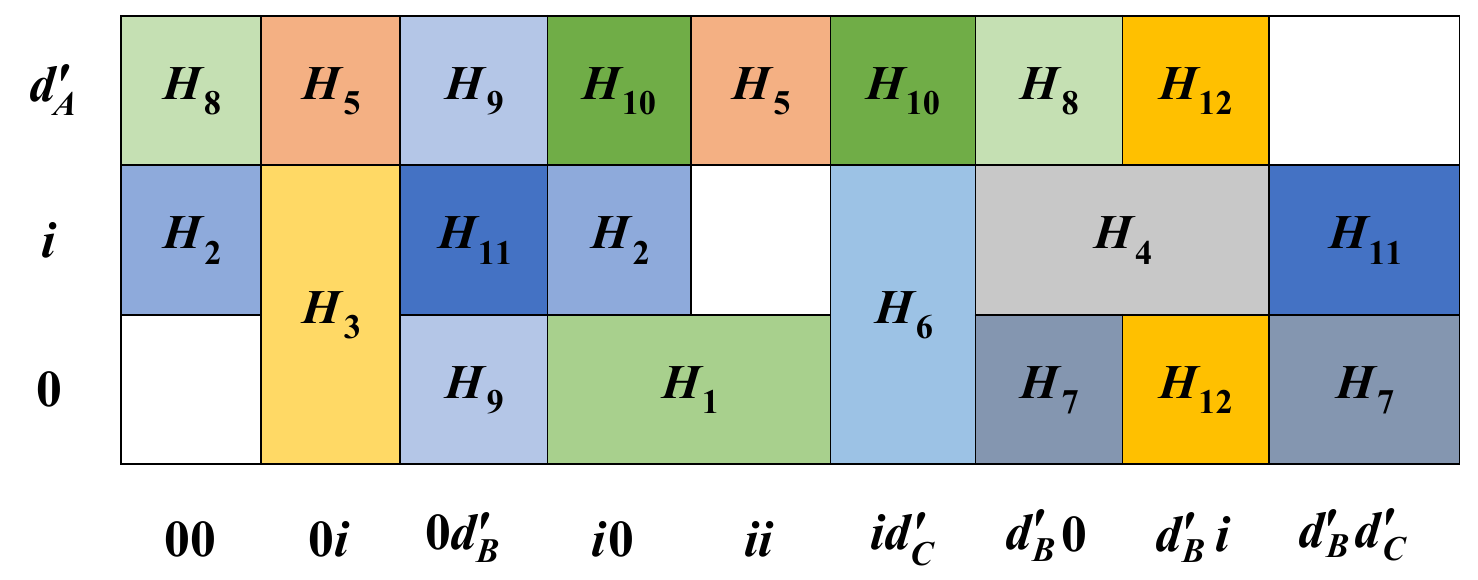}
\caption{The corresponding $d_{A}\times d_{B}d_{C}$ grid of $\{H_{i}\}_{i=1}^{12}$ given by Eq. (\ref{21}) in $A|BC$ bipartition. \label{10}}
\end{figure}

By extending the dimension of the grid in Fig. \ref{9}, we can generalize the structure of the set (\ref{20}) to any finite dimension. The OPS in $\mathcal{C}^{d_{A}}\otimes\mathcal{C}^{d_{B}}\otimes\mathcal{C}^{d_{C}}$ is described as
\begin{equation}\label{21}
\begin{aligned}
&H_{1}=\{|0\rangle_{A}|\xi_{i}\rangle_{B}|\eta_{j}\rangle_{C}\}_{i,j}, \\
&H_{2}=\{|\xi_{i}\rangle_{A}|\eta_{j}\rangle_{B}|0\rangle_{C}\}_{i,j}, \\
&H_{3}=\{|\eta_{j}\rangle_{A}|0\rangle_{B}|\xi_{i}\rangle_{C}\}_{i,j},\\
&H_{4}=\{|\xi_{i}\rangle_{A}|d_{B}'\rangle_{B}|\eta_{j}\rangle_{C}\}_{i,j}, \\
&H_{5}=\{|d_{A}'\rangle_{A}|\eta_{j}\rangle_{B}|\xi_{i}\rangle_{C}\}_{i,j}, \\
&H_{6}=\{|\eta_{j}\rangle_{A}|\xi_{i}\rangle_{B}|d_{C}'\rangle_{C}\}_{i,j},\\
&H_{7}=\{|0\rangle_{A}|d_{B}'\rangle_{B}|0\pm d_{C}'\rangle_{C}\}, \\
&H_{8}=\{|d_{A}'\rangle_{A}|0\pm d_{B}'\rangle_{B}|0\rangle_{C}\}, \\
&H_{9}=\{|0\pm d_{A}'\rangle_{A}|0\rangle_{B}|d_{C}'\rangle_{C}\},\\
&H_{10}=\{|d_{A}'\rangle_{A}|\xi_{i}\rangle_{B}|0\pm d_{C}'\rangle_{C}\}_{i}, \\
&H_{11}=\{|\xi_{i}\rangle_{A}|0\pm d_{B}'\rangle_{B}|d_{C}'\rangle_{C}\}_{i}, \\
&H_{12}=\{|0\pm d_{A}'\rangle_{A}|d_{B}'\rangle_{B}|\xi_{i}\rangle_{C}\}_{i},
\end{aligned}
\end{equation}
where $|\xi_{i}\rangle_{\tau}=\sum_{u=0}^{d_{\tau}-3}\omega_{d_{\tau}-2}^{iu}|u+1\rangle$, $|\eta_{j}\rangle_{\tau}=\sum_{u=0}^{d_{\tau}-2}$ $\omega_{d_{\tau}-1}^{ju}|u\rangle$, $d_{\tau}'=d_{\tau}-1$ for $i\in \mathcal{Z}_{d_{\tau}-2}$, $j\in \mathcal{Z}_{d_{\tau}-1}$, and $\tau\in\{A,B,C\}$. Here and below we use the notation $\omega_{n}:=\textup{e}^{\frac{2\pi\textup{i}}{n}}$ for any positive integer $n$. Fig. \ref{10} is a geometric representation of this OPS in $A|BC$ bipartition. We explain the srong nonlocality of the OPS (\ref{21}) in the following theorem.

\emph{Theorem~4}. In $\mathcal{C}^{d_{A}}\otimes\mathcal{C}^{d_{B}}\otimes\mathcal{C}^{d_{C}}$, the set $\cup_{i=1}^{12}H_{i}$ given by Eq. (\ref{21}) is an OPS of the strongest nonlocality. The size of this set is $2[(d_{A}d_{B}+d_{B}d_{C}+d_{A}d_{C})-2(d_{A}+d_{B}+d_{C})+3]$.

$Proof.$ We only need to discuss the orthogonality-preserving POVM performed on $BC$ party. The tile structure is depicted in Fig. \ref{10}. Because the set $\cup_{i=1}^{12}H_{i}$ has the same structure as the set $\cup_{i=1}^{12}S_{i}$ given by Eq. (\ref{20}), the conditions i), ii) and iv) of Theorem 1 are obvious. Here $R_{1}=H_{5}\cup H_{10}$, $R_{2}=H_{8}\cup H_{10}$, $R_{3}=H_{5}$, $R_{4}=H_{8}\cup H_{12}$, $R_{5}=H_{1}\cup H_{3}$, $R_{6}=H_{10}$, $R_{7}=H_{4}\cup H_{11}$, $R_{8}=H_{2}\cup H_{4}$, $R_{9}=H_{11}$, $R_{10}=H_{2}\cup H_{6}$, $R_{11}=H_{7}\cup H_{9}$ and $R_{12}=H_{4}$. Now consider the condition iii).

It is not difficult to show that the set sequence
\begin{equation*}
\begin{aligned}
&G_{1}=H_{2}\cup H_{4}\cup H_{7}\cup H_{8}\cup H_{9}\cup H_{11}, \\
&G_{2}=H_{1}\cup H_{10}\cup H_{12}, \\
&G_{3}=H_{5}\cup H_{6}, \\
&G_{4}=H_{3},
\end{aligned}
\end{equation*}
satisfies 1) and 2). Here each subset contained in $G_{x}$ $(x=2,3,4)$ is a NIC subset.
For $H_{1}\subset G_{2}$, we find that there are $H_{2}\subset G_{1}$ and $H_{10}\subset R_{1}$ such that $H_{1}^{(BC)}\cap H_{2}^{(BC)}=H_{1}^{(BC)}\cap H_{10}^{(BC)}$.
For the subsets $H_{10}$, $H_{12}$, $H_{5}$, $H_{6}$ and $H_{3}$, there are $H_{2}=G_{1}\cap R_{10}$, $H_{4}=G_{1}\cap R_{12}$, $H_{1}=G_{2}\cap R_{5}$, $H_{10}=G_{2}\cap R_{6}$ and $H_{5}=G_{3}\cap R_{3}$, respectively. It follows that the condition iii) in Theorem 1 holds.

According to Theorem 1, the orthogonality-preserving POVM performed on $BC$ party can only be trivial. Therefore, the set $\cup_{i=1}^{12}H_{i}$ given by Eq. (\ref{21}) is of the strongest nonlocality.
\hfill $\square$

\begin{figure}[h]
\centering
\includegraphics[width=0.48\textwidth]{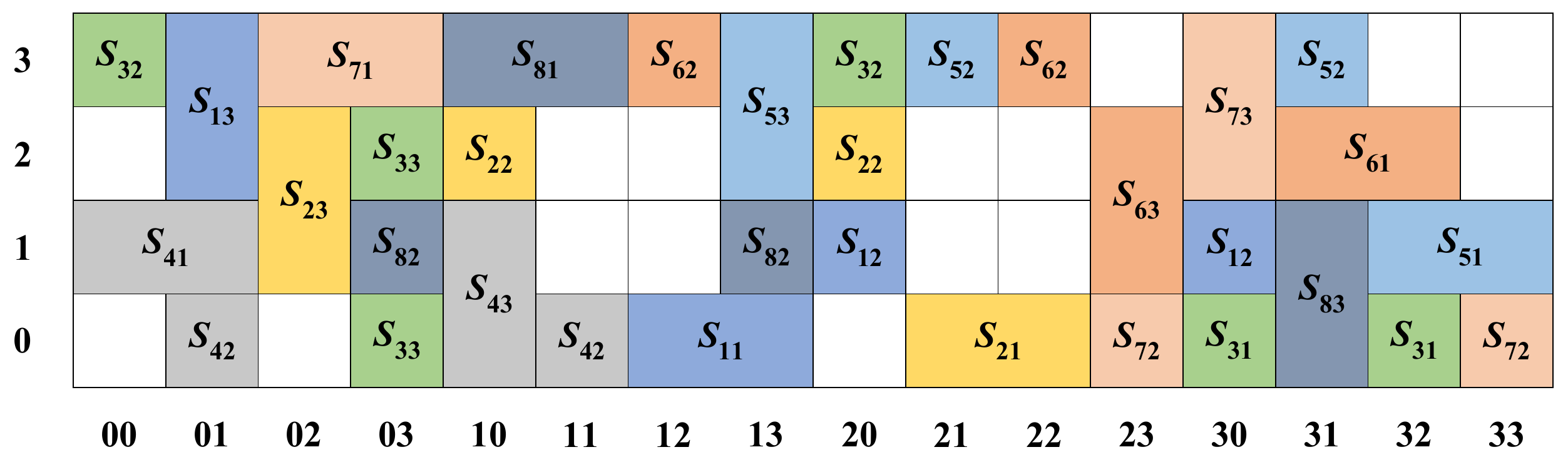}
\caption{The corresponding $4\times 16$ grid of $\{S_{ij}\}$ given by Eq. (\ref{22}) in $A|BC$ bipartition. \label{11}}
\end{figure}

Applying Theorem 1, we propose a set of strongly nonlocal OPS in  $\mathcal{C}^{4}\otimes\mathcal{C}^{4}\otimes\mathcal{C}^{4}$. The newly constructed OPS  contains fewer quantum states than in  Ref. \cite{Yuan, Shi2}. The specific OPS is given by
\begin{equation}\label{22}
\begin{aligned}
&S_{11}=\{|0\rangle|1\rangle|2\pm 3\rangle\}, && S_{51}=\{|1\rangle|3\rangle|2\pm 3\rangle\},\\
&S_{12}=\{|1\rangle|2\pm 3\rangle|0\rangle\}, && S_{52}=\{|3\rangle|2\pm 3\rangle|1\rangle\},\\
&S_{13}=\{|2\pm 3\rangle|0\rangle|1\rangle\}, && S_{53}=\{|2\pm 3\rangle|1\rangle|3\rangle\},\\
&S_{21}=\{|0\rangle|2\rangle|1\pm 2\rangle\}, && S_{61}=\{|2\rangle|3\rangle|1\pm 2\rangle\},\\
&S_{22}=\{|2\rangle|1\pm 2\rangle|0\rangle\}, && S_{62}=\{|3\rangle|1\pm 2\rangle|2\rangle\},\\
&S_{23}=\{|1\pm 2\rangle|0\rangle|2\rangle\}, && S_{63}=\{|1\pm 2\rangle|2\rangle|3\rangle\},\\
&S_{31}=\{|0\rangle|3\rangle|0\pm 2\rangle\}, && S_{71}=\{|3\rangle|0\rangle|2\pm 3\rangle\},\\
&S_{32}=\{|3\rangle|0\pm 2\rangle|0\rangle\}, && S_{72}=\{|0\rangle|2\pm 3\rangle|3\rangle\},\\
&S_{33}=\{|0\pm 2\rangle|0\rangle|3\rangle\}, && S_{73}=\{|2\pm 3\rangle|3\rangle|0\rangle\},\\
&S_{41}=\{|1\rangle|0\rangle|0\pm 1\rangle\}, && S_{81}=\{|3\rangle|1\rangle|0\pm 1\rangle\},\\
&S_{42}=\{|0\rangle|0\pm 1\rangle|1\rangle\}, && S_{82}=\{|1\rangle|0\pm 1\rangle|3\rangle\},\\
&S_{43}=\{|0\pm 1\rangle|1\rangle|0\rangle\}, && S_{83}=\{|0\pm 1\rangle|3\rangle|1\rangle\}.
\end{aligned}
\end{equation}
A geometric representation of this OPS in $A|BC$ bipartition is depicted in Fig. \ref{11}.

\emph{Theorem~5}. In $\mathcal{C}^{4}\otimes\mathcal{C}^{4}\otimes\mathcal{C}^{4}$, the set $\cup_{i=1}^{8}(\cup_{j=1}^{3}S_{ij})$ given by Eq. (\ref{22}) is of the strongest nonlocality. The size of this set is 48.

The detailed proof is shown in Appendix \ref{E}. Up to now, we have constructed a strongly nolocal OPS  containing 48 states in $\mathcal{C}^{4}\otimes\mathcal{C}^{4}\otimes\mathcal{C}^{4}$, which is 6 and 8 fewer than states presented in Ref. \cite{Yuan} and \cite{Shi2}, respectively.

Next, we generalize the structures of OPSs given by Eq. (\ref{22}) and Ref. \cite{Yuan} to  systems $\mathcal{C}^{d_{A}}\otimes\mathcal{C}^{d_{B}}\otimes\mathcal{C}^{d_{C}}$ and $\mathcal{C}^{d_{A}}\otimes\mathcal{C}^{d_{B}}\otimes\mathcal{C}^{d_{C}}\otimes\mathcal{C}^{d_{D}}$, respectively.

In quantum system $\mathcal{C}^{d_{A}}\otimes\mathcal{C}^{d_{B}}\otimes\mathcal{C}^{d_{C}}$ $(d_{A},d_{B},d_{C}\geq 4)$, consider the following OPS
\begin{equation}\label{23}
\begin{aligned}
&H_{11}=\{|0\rangle_{A}|1\rangle_{B}|\alpha_{3}^{l}\rangle_{C}\}_{l}, \\
&H_{12}=\{|1\rangle_{A}|\alpha_{3}^{l}\rangle_{B}|0\rangle_{C}\}_{l}, \\
&H_{13}=\{|\alpha_{3}^{l}\rangle_{A}|0\rangle_{B}|1\rangle_{C}\}_{l}, \\
&H_{21}=\{|0\rangle_{A}|\alpha^{i}\rangle_{B}|\alpha_{1}^{k}\rangle_{C}\}_{i,k}, \\ &H_{22}=\{|\alpha^{i}\rangle_{A}|\alpha_{1}^{k}\rangle_{B}|0\rangle_{C}\}_{i,k}, \\ &H_{23}=\{|\alpha_{1}^{k}\rangle_{A}|0\rangle_{B}|\alpha^{i}\rangle_{C}\}_{i,k}, \\
&H_{31}=\{|0\rangle_{A}|d_{B}'\rangle_{B}|\alpha_{0}^{j}\rangle_{C}\}_{j}, \\
&H_{32}=\{|d_{A}'\rangle_{A}|\alpha_{0}^{j}\rangle_{B}|0\rangle_{C}\}_{j}, \\
&H_{33}=\{|\alpha_{0}^{j}\rangle_{A}|0\rangle_{B}|d_{C}'\rangle_{C}\}_{j}, \\
&H_{41}=\{|1\rangle_{A}|0\rangle_{B}|0\pm 1\rangle_{C}\}, \\
&H_{42}=\{|0\rangle_{A}|0\pm 1\rangle_{B}|1\rangle_{C}\}, \\
&H_{43}=\{|0\pm 1\rangle_{A}|1\rangle_{B}|0\rangle_{C}\}, \\
&H_{51}=\{|1\rangle_{A}|d_{B}'\rangle_{B}|\alpha_{3}^{l}\rangle_{C}\}_{l}, \\ &H_{52}=\{|d_{A}'\rangle_{A}|\alpha_{3}^{l}\rangle_{B}|1\rangle_{C}\}_{l}, \\ &H_{53}=\{|\alpha_{3}^{l}\rangle_{A}|1\rangle_{B}|d_{C}'\rangle_{C}\}_{l}, \\
&H_{61}=\{|\alpha^{i}\rangle_{A}|d_{B}'\rangle_{B}|\alpha_{1}^{k}\rangle_{C}\}_{i,k}, \\
&H_{62}=\{|d_{A}'\rangle_{A}|\alpha_{1}^{k}\rangle_{B}|\alpha^{i}\rangle_{C}\}_{i,k}, \\ &H_{63}=\{|\alpha_{1}^{k}\rangle_{A}|\alpha^{i}\rangle_{B}|d_{C}'\rangle_{C}\}_{i,k}, \\
&H_{71}=\{|d_{A}'\rangle_{A}|0\rangle_{B}|\alpha_{3}^{l}\rangle_{C}\}_{l}, \\ &H_{72}=\{|0\rangle_{A}|\alpha_{3}^{l}\rangle_{B}|d_{C}'\rangle_{C}\}_{l}, \\ &H_{73}=\{|\alpha_{3}^{l}\rangle_{A}|d_{B}'\rangle_{B}|0\rangle_{C}\}_{l}, \\
&H_{81}=\{|d_{A}'\rangle_{A}|1\rangle_{B}|0\pm 1\rangle_{C}\}, \\
&H_{82}=\{|1\rangle_{A}|0\pm 1\rangle_{B}|d_{C}'\rangle_{C}\}, \\
&H_{83}=\{|0\pm 1\rangle_{A}|d_{B}'\rangle_{B}|1\rangle_{C}\}.
\end{aligned}
\end{equation}
Here $|\alpha^{i}\rangle_{\tau}=\sum_{u=0}^{d_{\tau}-4}\omega_{d_{\tau}-3}^{iu}|u+2\rangle$, $|\alpha_{0}^{j}\rangle_{\tau}=|0\rangle+\sum_{u=1}^{d_{\tau}-3}\omega_{d_{\tau}-2}^{ju}|u+1\rangle$, $|\alpha_{1}^{k}\rangle_{\tau}=\sum_{u=0}^{d_{\tau}-3}\omega_{d_{\tau}-2}^{ku}|u+1\rangle$, $|\alpha_{3}^{l}\rangle_{\tau}=\sum_{u=0}^{d_{\tau}-3}\omega_{d_{\tau}-2}^{lu}|u+2\rangle$, $d_{\tau}'=d_{\tau}-1$ for $i\in \mathcal{Z}_{d_{\tau}-3}$, $j,k,l\in \mathcal{Z}_{d_{\tau}-2}$ and $\tau=A,B,C$. Since the above OPS has the same structure as the set (\ref{22}), we find that it is  strong nonlocal.

\emph{Theorem~6}. In $\mathcal{C}^{d_{A}}\otimes\mathcal{C}^{d_{B}}\otimes\mathcal{C}^{d_{C}}$, the set $\cup_{i=1}^{8}(\cup_{j=1}^{3}H_{ij})$ given by Eq. (\ref{23}) is an OPS of the strongest nonlocality. The size of this set is $2[(d_{A}d_{B}+d_{B}d_{C}+d_{A}d_{C})-3(d_{A}+d_{B}+d_{C})+12]$.

The detailed proof is in Appendix \ref{F}.  In $\mathcal{C}^{d}\otimes\mathcal{C}^{d}\otimes\mathcal{C}^{d}$, the  size $6[(d-1)^2-d+3]$ of the strongly nonlocal OPS of Theorem 4  is strictly fewer, $6(d-3)$ fewer to be precise,  than the size $6(d-1)^{2}$ of the strongly nonlocal OPS in Ref. \cite{Yuan}.
Similarly, we propose the following OPS in $\mathcal{C}^{d_{A}}\otimes\mathcal{C}^{d_{B}}\otimes\mathcal{C}^{d_{C}}\otimes\mathcal{C}^{d_{D}}$
\begin{equation}\label{24}
\begin{aligned}
&U_{11}=\{|0\rangle_{A}|\xi_{i}\rangle_{B}|\eta_{j}\rangle_{C}|0\pm d_{D}'\rangle_{D}\}_{i,j}, \\
&U_{12}=\{|\xi_{i}\rangle_{A}|\eta_{j}\rangle_{B}|0\pm d_{C}'\rangle_{C}|0\rangle_{D}\}_{i,j}, \\
&U_{13}=\{|\eta_{j}\rangle_{A}|0\pm d_{B}'\rangle_{B}|0\rangle_{C}|\xi_{i}\rangle_{D}\}_{i,j}, \\
&U_{14}=\{|0\pm d_{A}'\rangle_{A}|0\rangle_{B}|\xi_{i}\rangle_{C}|\eta_{j}\rangle_{D}\}_{i,j}, \\
&U_{21}=\{|\xi_{i}\rangle_{A}|d_{B}'\rangle_{B}|\gamma_{k}\rangle_{C}|\eta_{j}\rangle_{D}\}_{i,j,k}, \\
&U_{22}=\{|d_{A}'\rangle_{A}|\gamma_{k}\rangle_{B}|\eta_{j}\rangle_{C}|\xi_{i}\rangle_{D}\}_{i,j,k}, \\
&U_{23}=\{|\gamma_{k}\rangle_{A}|\eta_{j}\rangle_{B}|\xi_{i}\rangle_{C}|d_{D}'\rangle_{D}\}_{i,j,k}, \\
&U_{24}=\{|\eta_{j}\rangle_{A}|\xi_{i}\rangle_{B}|d_{C}'\rangle_{C}|\gamma_{k}\rangle_{D}\}_{i,j,k}, \\
&U_{31}=\{|d_{A}'\rangle_{A}|0\rangle_{B}|0\pm d_{C}'\rangle_{C}|\gamma_{k}\rangle_{D}\}_{k}, \\
&U_{32}=\{|0\rangle_{A}|0\pm d_{B}'\rangle_{B}|\gamma_{k}\rangle_{C}|d_{D}'\rangle_{D}\}_{k}, \\
&U_{33}=\{|0\pm d_{A}'\rangle_{A}|\gamma_{k}\rangle_{B}|d_{C}'\rangle_{C}|0\rangle_{D}\}_{k}, \\
&U_{34}=\{|\gamma_{k}\rangle_{A}|d_{B}'\rangle_{B}|0\rangle_{C}|0\pm d_{D}'\rangle_{D}\}_{k}, \\
&U_{41}=\{|\xi_{i}\rangle_{A}|\xi_{i}\rangle_{B}|0\rangle_{C}|\gamma_{k}\rangle_{D}\}_{i|_{A},i|_{B},k}, \\
&U_{42}=\{|\xi_{i}\rangle_{A}|0\rangle_{B}|\gamma_{k}\rangle_{C}|\xi_{i}\rangle_{D}\}_{i|_{A},i|_{D},k}, \\
&U_{43}=\{|0\rangle_{A}|\gamma_{k}\rangle_{B}|\xi_{i}\rangle_{C}|\xi_{i}\rangle_{D}\}_{i|_{C},i|_{D},k}, \\
&U_{44}=\{|\gamma_{k}\rangle_{A}|\xi_{i}\rangle_{B}|\xi_{i}\rangle_{C}|0\rangle_{D}\}_{i|_{B},i|_{C},k}, \\
&U_{51}=\{|d_{A}'\rangle_{A}|d_{B}'\rangle_{B}|\xi_{i}\rangle_{C}|0\pm d_{D}'\rangle_{D}\}_{i}, \\
&U_{52}=\{|d_{A}'\rangle_{A}|\xi_{i}\rangle_{B}|0\pm d_{C}'\rangle_{C}|d_{D}'\rangle_{D}\}_{i}, \\
&U_{53}=\{|\xi_{i}\rangle_{A}|0\pm d_{B}'\rangle_{B}|d_{C}'\rangle_{C}|d_{D}'\rangle_{D}\}_{i}, \\
&U_{54}=\{|0\pm d_{A}'\rangle_{A}|d_{B}'\rangle_{B}|d_{C}'\rangle_{C}|\xi_{i}\rangle_{D}\}_{i}, \\
&U_{61}=\{|0\rangle_{A}|0\rangle_{B}|d_{C}'\rangle_{C}|\eta_{j}\rangle_{D}\}_{j}, \\
&U_{62}=\{|0\rangle_{A}|d_{B}'\rangle_{B}|\eta_{j}\rangle_{C}|0\rangle_{D}\}_{j}, \\
&U_{63}=\{|d_{A}'\rangle_{A}|\eta_{j}\rangle_{B}|0\rangle_{C}|0\rangle_{D}\}_{j}, \\
&U_{64}=\{|\eta_{j}\rangle_{A}|0\rangle_{B}|0\rangle_{C}|d_{D}'\rangle_{D}\}_{j}, \\
&U_{71}=\{|0\rangle_{A}|\xi_{i}\rangle_{B}|0\rangle_{C}|\xi_{i}\rangle_{D}\}_{i|_{B},i|_{D}}, \\
&U_{72}=\{|\xi_{i}\rangle_{A}|0\rangle_{B}|\xi_{i}\rangle_{C}|0\rangle_{D}\}_{i|_{A},i|_{C}}, \\
&U_{81}=\{|0\rangle_{A}|d_{B}'\rangle_{B}|0\rangle_{C}|d_{D}'\rangle_{D}\}, \\
&U_{82}=\{|d_{A}'\rangle_{A}|0\rangle_{B}|d_{C}'\rangle_{C}|0\rangle_{D}\}, \\
&U_{91}=\{|\xi_{i}\rangle_{A}|d_{B}'\rangle_{B}|\xi_{i}\rangle_{C}|d_{D}'\rangle_{D}\}_{i|_{A},i|_{C}}, \\
&U_{92}=\{|d_{A}'\rangle_{A}|\xi_{i}\rangle_{B}|d_{C}'\rangle_{C}|\xi_{i}\rangle_{D}\}_{i|_{B},i|_{D}},
\end{aligned}
\end{equation}
where $|\xi_{i}\rangle_{\tau}=\sum_{u=0}^{d_{\tau}-3}\omega_{d_{\tau}-2}^{iu}|u+1\rangle$, $|\eta_{j}\rangle_{\tau}=\sum_{u=0}^{d_{\tau}-2}$ $\omega_{d_{\tau}-1}^{ju}|u\rangle$, $|\gamma_{k}\rangle_{\tau}=\sum_{u=0}^{d_{\tau}-2}\omega_{d_{\tau}-1}^{ku}|u+1\rangle$, $d_{\tau}'=d_{\tau}-1$ for $i\in \mathcal{Z}_{d_{\tau}-2}$, $j,k\in \mathcal{Z}_{d_{\tau}-1}$, and $\tau=A,B,C,D$.

\emph{Theorem~7}. In the system $\mathcal{C}^{d_{A}}\otimes\mathcal{C}^{d_{B}}\otimes\mathcal{C}^{d_{C}}\otimes\mathcal{C}^{d_{D}}$, the set $\{\cup_{i=1}^{6}(\cup_{j=1}^{4}U_{ij})\}\cup\{\cup_{i=7}^{9}(\cup_{j=1}^{2}U_{ij})\}$ given by Eq. (\ref{24}) is an OPS of the strongest nonlocality. The size of this set is $d_{A}d_{B}d_{C}d_{D}-(d_{A}-2)(d_{B}-2)(d_{C}-2)(d_{D}-2)-2$.

The detailed proof is shown in Appendix \ref{G}. It is worth noting that the set (\ref{24}) is still of the strongest nonlocality even though it contains fewer quantum states than the set in Ref. \cite{Yuan}. Moreover, its size is smaller than that of the strongly nonlocal OPS in Ref. \cite{Shi2}.

Each of Theorems 2-7 gives a positive answer to one open problem in Ref. \cite{Halder} of ``whether incomplete orthogonal product bases can be strongly nonlocal.''

\section{entanglement-assisted discrimination}\label{Q5}

The above OPSs cannot be distinguished under LOCC even if any $n-1$ parties are allowed to come together. However, it is possible while one equips enough entanglement resource. Let  $|\phi^{+}(d)\rangle$ denote the  maximally entangled state $\frac{1}{\sqrt{d}}\sum_{i=0}^{d-1}|ii\rangle$ in $\mathcal{C}^{d}\otimes \mathcal{C}^{d}$. Let $(s,|\phi^{+}(d)\rangle_{AB})$ express a resource configuration, which means that on average an amount $s$ of the two-qudit maximally entangled state is consumed between Alice and Bob. In this section, we will present several different entanglement-assisted discrimination protocols. Without loss of generality, from now, we only consider in the case $d_{A}\geq d_{B}\geq d_{C}\geq d_{D}$.

\emph{Theorem~8}. The entanglement resource configuration $\{(1,|\phi^{+}(2)\rangle_{AB});(1,|\phi^{+}(d_{C})\rangle_{BC})\}$ is sufficient for local discrimination of the set ({\ref{21}}).

The detailed process is provided in Appendix \ref{H}. In this protocol, we use quantum teleportation one time and consume $(1+\log_{2} d_{C})$-ebit entanglement resource in total. It is strictly less than the amount consumed in the protocol which teleports all subsystems to one party. Next, we discuss the local discrimination of OPS ({\ref{21}}) without teleportation.

\emph{Theorem~9}. When all the parties are separated, the set $\cup_{i=1}^{12}H_{i}$ given by Eq.~({\ref{21}}) can be locally distinguished by using the entanglement resource $\{(s,|\phi^{+}(2)\rangle_{AB});$ $(1,|\phi^{+}(2)\rangle_{AC})\}$, where $s=1+\frac{e-3f+6}{2e-4f+6}$ for $e=d_{A}d_{B}+d_{A}d_{C}+d_{B}d_{C}$ and $f=d_{A}+d_{B}+d_{C}$.

The specific process is given in Appendix \ref{I}. The entanglement consumed in this protocol is $(1+s)$-ebit, due to $s<1.5<\log_{2} d_{C}$, which is less than the resource used in Theorem 8. Since the set (\ref{22}) is a special case of (\ref{23}) and they have the same structure, we only need to consider the entanglement-assisted discrimination protocols for the set ({\ref{23}}).

\emph{Theorem~10}. The set $\cup_{i=1}^{8}(\cup_{j=1}^{3}H_{ij})$ given by Eq. ({\ref{23}}) can be locally distinguished by using the entanglement resource configuration $\{(1,|\phi^{+}(2)\rangle_{AB});(1,|\phi^{+}(d_{C})\rangle_{BC})\}$.

\emph{Theorem~11}. The set $\cup_{i=1}^{8}(\cup_{j=1}^{3}H_{ij})$ given by Eq. ({\ref{23}}) can be locally distinguished by using the entanglement resource configuration $\{(1,|\phi^{+}(4)\rangle_{AB});(1,|\phi^{+}(2)\rangle_{AC})\}$.

The detailed proofs of Theorems 10 and 11 are given in Appendix \ref{J} and \ref{K}, respectively. The protocol in Theorem 10 uses teleportation while the protocol in Theorem 11 does not. Clearly $1+\log_{2} d_{C}$ ebits entanglement is consumed in the previous protocol, which is not less than the amount used 3 ebits in the latter protocol because $d_{C}\geq 4$. In other word, the latter resource configuration is more effective when the smallest dimension $d_{C}$ is greater than 4. Next, by the method presented by Zhang et al. in Ref. \cite{Zhang5}, using multiple copies of EPR states instead of high-dimensional entangled states, we can get a new resource configuration.

\emph{Theorem~12}. The entanglement resource configuration $\{(2,|\phi^{+}(2)\rangle_{AB});(1,|\phi^{+}(2)\rangle_{AC})\}$ is sufficient for local discrimination of the set $\cup_{i=1}^{8}(\cup_{j=1}^{3}H_{ij})$ given by Eq. ({\ref{23}}).

In fact, using two EPR states has the same effect as using one maximally entangled state $|\phi^{+}(4)\rangle_{AB}$. In the ancillary system of one party, $|00\rangle$, $|01\rangle$, $|10\rangle$ and $|11\rangle$ can correspond to $|0\rangle$, $|1\rangle$, $|2\rangle$ and $|3\rangle$, respectively. For the detailed procedure please refer to Appendix \ref{L}. This also shows that, in the similar discrimination protocol, we can replace a maximally entangled state $|\phi^{+}(d)\rangle$ with $n$ EPR states when $2^{n}\geq d$. Although more resources may be used, the method should be relatively easier to implement in real experiment because it only requires a device which can produce 2-qubit maximally entangled states. Besides, we also get several entanglement resource configurations to discriminate the set ({\ref{24}}) by LOCC.

\emph{Theorem~13}. The entanglement resource configuration $\{(1,|\phi^{+}(3)\rangle_{AB});(1,|\phi^{+}(d_{C})\rangle_{BC});(1,|\phi^{+}(d_{D})\rangle_{BD})\}$ is sufficient for local discrimination of the set $\{\cup_{i=1}^{6}(\cup_{j=1}^{4}U_{ij})\}\cup\{\cup_{i=7}^{9}(\cup_{j=1}^{2}U_{ij})\}$ given by Eq. ({\ref{24}}).

The protocol of Theorem 13 is given in Appendix \ref{M}.

\emph{Theorem~14}. Any one of the resource configurations $\{(1,|\phi^{+}(3)\rangle_{AB});(1,|\phi^{+}(3)\rangle_{AC});(1,|\phi^{+}(3)\rangle_{AD})\}$ and $\{(2,|\phi^{+}(2)\rangle_{AB});(2,|\phi^{+}(2)\rangle_{AC});(2,|\phi^{+}(2)\rangle_{AD})\}$ is sufficient for local discrimination of the set ({\ref{24}}).

We will not repeat the protocol of Theorem 14, because it is similar to that of Theorems 11 and 12. In Theorem 13, we perform quantum teleportation twice and consume $\log_{2} 3d_{C}d_{D}$ ebits entanglement resource. In comparison, the first configuration of Theorem 14 is more effective because $\log_{2} 27\leq \log_{2} 3d_{C}d_{D}$, and the second configuration is simpler because it only needs multiple EPR states.

\section{Conclusion}\label{Q6}
We have investigated the OPS with strong quantum nonlocality in multipartite quantum systems through the decomposition of plane geometry. Sufficient conditions for the triviality of orthogonality-preserving POVM on fixed subsystem are presented.
We have shown the minimum size of strongly unlocal OPSs  under some restrictions in $\mathcal{C}^{3}\otimes \mathcal{C}^{3}\otimes \mathcal{C}^{3}$ and $\mathcal{C}^{4}\otimes \mathcal{C}^{4}\otimes \mathcal{C}^{4}$, which partially answer an open question in Ref.\cite{Yuan}:  ``Can we find the smallest strongly nonlocal set in $\mathcal{C}^{3}\otimes\mathcal{C}^{3}\otimes\mathcal{C}^{3}$, and more generally in any tripartite systems?''.
Furthermore, we successfully constructed a smaller OPS which has the strongest nonlocality in $\mathcal{C}^{d_{A}}\otimes \mathcal{C}^{d_{B}}\otimes \mathcal{C}^{d_{C}}$ $(d_{A},d_{B},d_{C}\geq 4)$ and generalized the previous known structures of strongly nonlocal OPSs to any possible three and four-partite systems. Interestingly, we studied local discrimination protocols for our OPSs with different types of entangled resources. Among them, we have three protocols which only need multiple copies of EPR states. We found that the protocols without teleportation can be more efficient on average. More than that, our results could also be helpful in better understanding of the properties of maximally entangled states.

\begin{acknowledgments}
This work was supported by the National Natural Science Foundation of China under Grant Nos. 12071110 and 62271189, the Hebei Natural Science Foundation of China under Grant No. A2020205014, the Science and Technology Project of Hebei Education Department under Grant Nos. ZD2020167 and ZD2021066, and funded by School of Mathematical Sciences of Hebei Normal University under Grant No. 2021sxbs002.
\end{acknowledgments}

\begin{appendix}
\section{The proof of theorem 2}\label{B}
According to Corollary 2, we know the union $\cup_{r}S_{r}^{(BC)}$ of all projection sets is the basis $\mathcal{B}^{BC}$ and the family of projection sets $\{S_{r}^{(BC)}\}_{r}$ is connected.

When $N=1$, it is obvious that the set $S$ is locally distinguishable. When $N=2$, due to the symmetry, there is the collection $\{S_{t_{1}},S_{t_{2}},S_{t_{3}}\}$ including 6 quantum states, which satisfies $|S_{t_{1}}|=|S_{t_{2}}|=|S_{t_{3}}|=2$, $|S_{t_{1}}^{(BC)}|=|S_{t_{2}}^{(BC)}|=2$ and $|S_{t_{3}}^{(BC)}|=1$. Moreover, the collection is invariant under the cyclic permutation of the parties. According to the completeness and connectedness of projection sets, the set $S$ contains at least 8 subsets whose projection sets on $BC$ party have two elements. That is, we have no less than 4 disjoint collections with above form. In other words, when $N=2$, the size of set $S$ cannot be less than 24.

The case $N=3$ does not exist. If $N=3$, then there must be a subset satisfying $|S_{t}^{(A)}|=3$ and $|S_{t}^{(BC)}|=1$. Meanwhile, $S_{t}^{(A)}=\mathcal{B}^A$. We have $S_{t}^{(BC)}\cap (\cup_{t'\in Q\setminus\{t\}}S_{t'}^{(BC)})=\emptyset$. Hence, the family of projection sets $\{S_{r}^{(BC)}\}_{r}$ is unconnected, which is contradiction.  Similarly, the cases $N=6,9$ do not exist.

In the case $N=4$, because of symmetry, there is a collection $\{S_{u_{1}},S_{u_{2}},S_{u_{3}}\}$ containing 12 quantum states, which is symmetric and satisfies $|S_{u_{1}}|=|S_{u_{2}}|=|S_{u_{3}}|=4$, $|S_{u_{1}}^{(BC)}|=4$ and $|S_{u_{2}}^{(BC)}|=|S_{u_{3}}^{(BC)}|=2$. Similarly, due to the completeness and connectedness of projection sets, there are at least another subset whose projection set on $BC$ party has four elements or three additional subsets whose projection sets on $BC$ party have two elements. In either case, it means that the size of set $S$ is not less than 24.

It is obvious that $|S_{r}|\neq 5,7$ for any $r\in Q$. If there is a subset such that $|S_{r}|=8$, then for arbitrary cyclic permutation $P_{c}$ of subsystems, the two subspaces spanned by $S_{r}$ and $P_{c}(S_{r})$, respectively, are not orthogonal. It follows that there must be two nonorthogonal quantum states, one of which  belongs to $S_{r}$ and the other of which belongs to $P_{c}(S_{r})$. This contradicts the fact $P_{c}(S_{r})\subset S$. Consequently, the cases $N=5,7,8$ do not hold.

On the other hand, the strongly nonlocal OPS given by Eq. (\ref{20}) satisfies all conditions and contains 24 quantum states. Thus, in $\mathcal{C}^{3}\otimes\mathcal{C}^{3}\otimes\mathcal{C}^{3}$, the minimum size of the set $S$ is 24. The proof is completed.

\section{The proof of theorem 3}\label{C}
Because the set is symmetric and the maximum size of all subsets is 2, there is a collection $\{S_{t_{1}},S_{t_{2}},S_{t_{3}}\}$ containing 6 quantum states. It satisfies the same requirements as the proof of Theorem 2. Due to the completeness and connectedness of projection sets, there are at least 15 subsets whose projection sets on $BC$ party have size 2. So, we have no less than 8 disjoint collections, each of which contains 6 quantum states. That is, the set $S$ contains at least 48 quantum states. On the other side, we find the OPS given by Eq. (\ref{22}) satisfies all conditions and the size is 48. Therefore, the minimum size of set $S$ is 48.

\section{The proof of theorem 5}\label{E}
According to Lemma 1 and the invariance of the set (\ref{22}) under cyclic permutations, we only need to discuss the orthogonal-preserving measurement on $BC$ party. The tile structure is illustrated in Fig. \ref{11}. It is obvious $\tilde{S}_{V_{kl}}=\mathcal{B}^{BC}$ for all $k,l\in\mathcal{Z}_{4}$, which implies that $\mathcal{B}_{kl}^{BC}\subset\tilde{S}_{V_{kl}}$. Hence the condition i) holds.

For each subset $S_{ij}$, there is the corresponding PI set $R_{ij}$, which is shown in table \ref{30}.  It follows that the condition ii) is satisfied.
\begin{table}[tbp]
\centering
\caption{Corresponding PI set $R_{ij}$ for each subset $S_{ij}$.}\label{30}
\begin{tabular}{cl|cl}
\hline
\hline
Subset~~ & ~~~~~PI set~~~~~ & ~~Subset~~ & ~~~~~PI set~~~~~ \\ \hline
 $S_{11}$ & $R_{11}=S_{53}\cup S_{62}$ & $S_{51}$ & $R_{51}=S_{31}\cup S_{72}$ \\
 $S_{12}$ & $R_{12}=S_{32}\cup S_{73}$ & $S_{52}$ & $R_{52}=S_{21}\cup S_{83}$ \\
 $S_{13}$ & $R_{13}=S_{41}$            & $S_{53}$ & $R_{53}=S_{11}$            \\
 $S_{21}$ & $R_{21}=S_{52}\cup S_{62}$ & $S_{61}$ & $R_{61}=S_{51}\cup S_{83}$ \\
 $S_{22}$ & $R_{22}=S_{32}\cup S_{81}$ & $S_{62}$ & $R_{62}=S_{11}\cup S_{21}$ \\
 $S_{23}$ & $R_{23}=S_{71}$            & $S_{63}$ & $R_{63}=S_{72}$            \\
 $S_{31}$ & $R_{31}=S_{12}\cup S_{51}$ & $S_{71}$ & $R_{71}=S_{23}\cup S_{82}$ \\
 $S_{32}$ & $R_{32}=S_{12}\cup S_{41}$ & $S_{72}$ & $R_{72}=S_{51}\cup S_{63}$ \\
 $S_{33}$ & $R_{33}=S_{82}$            & $S_{73}$ & $R_{73}=S_{12}$            \\
 $S_{41}$ & $R_{41}=S_{13}\cup S_{32}$ & $S_{81}$ & $R_{81}=S_{42}\cup S_{43}$ \\
 $S_{42}$ & $R_{42}=S_{13}\cup S_{81}$ & $S_{82}$ & $R_{82}=S_{11}\cup S_{33}$ \\
 $S_{43}$ & $R_{43}=S_{81}$            & $S_{83}$ & $R_{83}=S_{61}$            \\
\hline
\end{tabular}
\end{table}

Since $|S_{ij}^{(BC)}\cap S_{kl}^{(BC)}|\leq 1$ for any two subsets $S_{ij}$ and $S_{kl}$, each $R_{ij}$ is a UPI set. Therefore $G_{1}$ is the union of all subsets. Thus, the condition iii) is true.

In addition, we have a sequence of projection sets $S_{41}^{(BC)}\rightarrow S_{42}^{(BC)}\rightarrow S_{81}^{(BC)}\rightarrow S_{22}^{(BC)}\rightarrow S_{12}^{(BC)}\rightarrow S_{31}^{(BC)}\rightarrow S_{51}^{(BC)}~(\rightarrow S_{72}^{(BC)})\rightarrow S_{61}^{(BC)}\rightarrow S_{52}^{(BC)}\rightarrow S_{21}^{(BC)}\rightarrow S_{62}^{(BC)}\rightarrow S_{11}^{(BC)}\rightarrow S_{82}^{(BC)}\rightarrow S_{71}^{(BC)}$, where the intersection of the sets on both sides of the arrow is not empty and the union of these sets is the computation basis $\mathcal{B}^{BC}$. Here the set $S_{72}^{(BC)}$ in the bracket is only related to the previous set $S_{51}^{(BC)}$. This means that it is impossible to divide all projection sets into disjoint two groups. That is, the family of projection sets $\{S_{ij}^{(BC)}\}_{ij}$ is connected. The condition iv) holds.

By using Theorem 1, the orthogonality-preserving POVM performed on $BC$ party can only be trivial. Therefore, the OPS (\ref{22}) is of the strongest quantum nonlocality.

\section{The proof of theorem 6}\label{F}
We need only to consider the orthogonality-preserving POVM on $BC$ party. Because the set (\ref{23}) has the same structure as the set (\ref{22}), the conditions i), ii) and iv) are obvious. Given the set sequence
\begin{equation}
\begin{aligned}
G_{1}=& H_{11}\cup H_{12}\cup H_{13}\cup H_{22}\cup H_{31}\cup H_{32}\cup H_{33}\\
&\cup H_{41}\cup H_{42}\cup H_{43}\cup H_{51}\cup H_{52}\cup H_{53}\cup H_{61}\\
&\cup H_{71}\cup H_{72}\cup H_{73}\cup H_{81}\cup H_{82}\cup H_{83}, \\
G_{2}=& H_{21}\cup H_{23}\cup H_{62}\cup H_{63}.
\end{aligned}
\end{equation}
Here each subset contained in $G_{2}$ is a NIC subset.
Referring to table \ref{30}, we can get the PI set $R_{ij}$ of $H_{ij}$ on $BC$ party. More specifically,  $H_{ij}$ is substituted for $S_{ij}$ in table \ref{30}, one gets the PI set $R_{ij}$ of $H_{ij}$ on $BC$ party.   For the subsets $H_{21}$, $H_{23}$, $H_{62}$ and $H_{63}$, there are $H_{52}=G_{1}\cap R_{21}$, $H_{71}=G_{1}\cap R_{23}$, $H_{11}=G_{1}\cap R_{62}$ and $H_{72}=G_{1}\cap R_{63}$, respectively. It implies that condition iii) holds.

The set (\ref{23}) satisfies the four conditions in Theorem 1, therefore it is locally irreducible in every bipartition. That is, the OPS (\ref{23}) is a set of the strongest nonlocality.

\section{The proof of theorem 7}\label{G}
\begin{figure}[h]
\centering
\includegraphics[width=0.48\textwidth]{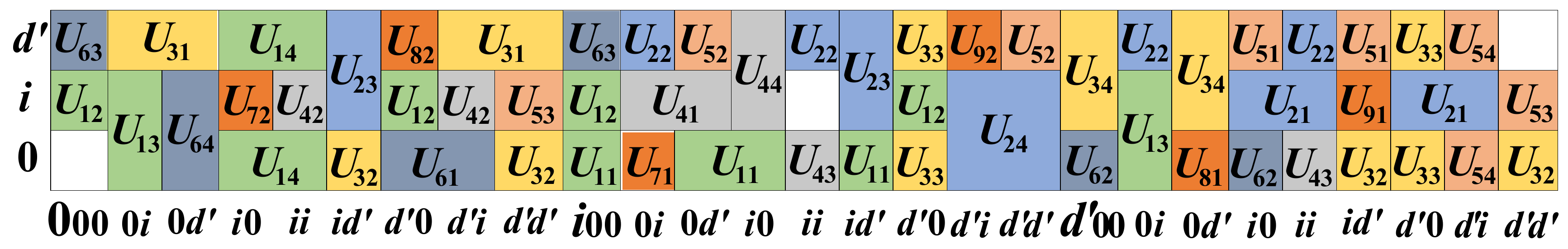}
\caption{The corresponding $3\times 27$ grid of $\{U_{ij}\}$ given by Eq. (\ref{24}) in $A|BCD$ bipartition. \label{12}}
\end{figure}

We need to prove that the orthogonality-preserving POVM performed on $BCD$ party can only be trivial. To see this, we will prove that the OPS (\ref{24}) satisfies the four conditions in Theorem 1.

Fig. \ref{12} is the tile structure of this OPS. Note that $\tilde{S}_{V_{jkl}}=\mathcal{B}^{BCD}$ for any $j,k,l\in\mathcal{Z}_{3}$. It is obvious $\mathcal{B}_{jkl}^{BCD}\subset\tilde{S}_{V_{jkl}}$. The condition i) holds.

For each subset $U_{ij}$, there is the corresponding PI set $R_{ij}$, which is shown in table \ref{32}. Hence, the condition ii) holds.
\begin{table}[tbp]
\centering
\caption{Corresponding PI set $R_{ij}$ for each subset $U_{ij}$.}\label{32}
\begin{tabular}{cl|cl}
\hline
\hline
Subset~~ & ~~~~~~~~~~~PI set~~~~~~~~~~~ &~~Subset~~ & ~~~~~~~~PI set~~~~~~~~~~~ \\ \hline
 $U_{11}$ & \tabincell{l}{$R_{11}=U_{12}\cup U_{23}$\\$~~~~~~~~\cup U_{41}\cup U_{44}$} &  $U_{44}$ & $R_{44}=U_{11}$            \\ \hline
 $U_{12}$ & \tabincell{l}{$R_{12}=U_{33}\cup U_{63}$\\$~~~~~~~~\cup U_{82}$}            &  $U_{51}$ & $R_{51}=U_{32}\cup U_{62}$ \\ \hline
 $U_{13}$ & $R_{13}=U_{22}\cup U_{31}$                       &  $U_{52}$ & $R_{52}=U_{11}\cup U_{24}$ \\ \hline
 $U_{14}$ & $R_{14}=U_{42}\cup U_{72}$                       &  $U_{53}$ & $R_{53}=U_{32}$            \\ \hline
 $U_{21}$ & \tabincell{l}{$R_{21}=U_{22}\cup U_{33}$\\$~~~~~~~~\cup U_{51}\cup U_{54}$} &  $U_{54}$ & $R_{54}=U_{21}$            \\ \hline
 $U_{22}$ & \tabincell{l}{$R_{22}=U_{13}\cup U_{43}$\\$~~~~~~~~\cup U_{71}$}            &  $U_{61}$ & $R_{61}=U_{12}\cup U_{42}$\\\hline
 $U_{23}$ & $R_{23}=U_{11}\cup U_{32}$                       &  $U_{62}$ & $R_{62}=U_{34}\cup U_{51}$\\ \hline
 $U_{24}$ & $R_{24}=U_{52}\cup U_{92}$                       &  $U_{63}$ & $R_{63}=U_{12}$ \\ \hline
 $U_{31}$ & \tabincell{l}{$R_{31}=U_{13}\cup U_{42}$\\$~~~~~~~~\cup U_{53}\cup U_{64}$} &  $U_{64}$ & $R_{64}=U_{31}$ \\ \hline
 $U_{32}$ & \tabincell{l}{$R_{32}=U_{23}\cup U_{53}$\\$~~~~~~~~\cup U_{91}$}            &  $U_{71}$ & $R_{71}=U_{41}$ \\ \hline
 $U_{33}$ & $R_{33}=U_{12}\cup U_{21}$ & $U_{72}$ & $R_{72}=U_{14}$ \\ \hline
 $U_{34}$ & $R_{34}=U_{62}\cup U_{81}$ &  $U_{81}$ & $R_{81}=U_{34}$ \\ \hline
 $U_{41}$ & $R_{41}=U_{11}\cup U_{71}$ & $U_{82}$ & $R_{82}=U_{12}$ \\ \hline
 $U_{42}$ & $R_{42}=U_{14}\cup U_{61}$ & $U_{91}$ & $R_{91}=U_{51}$ \\ \hline
 $U_{43}$ & $R_{43}=U_{22}$            & $U_{92}$ & $R_{92}=U_{24}$ \\
\hline
\end{tabular}
\end{table}

Furthermore, we construct the set sequence
\begin{equation}
\begin{aligned}
&G_{1}=U_{12}\cup U_{21}\cup U_{31}\cup U_{33}\cup U_{34}\cup U_{61}\\
&\quad~~~~~\cup U_{62}\cup U_{64}\cup U_{81}\cup U_{82},\\
&G_{2}=U_{11}\cup U_{13}\cup U_{42}\cup U_{51}\cup U_{54}\cup U_{63},\\
&G_{3}=U_{14}\cup U_{22}\cup U_{23}\cup U_{32}\cup U_{41}\cup U_{44}\\
&\quad~~~~~\cup U_{52}\cup U_{91},\\
&G_{4}=U_{24}\cup U_{43}\cup U_{53}\cup U_{71}\cup U_{72},\\
&G_{5}=U_{92}.
\end{aligned}
\end{equation}
For the subset $U_{32}\subset G_{3}$, there are subsets $U_{51}\subset G_{2}$ and $U_{91}\subset R_{32}$ such that $U_{32}^{(BCD)}\cap U_{51}^{(BCD)}=U_{32}^{(BCD)}\cap U_{91}^{(BCD)}$. For any other subset $U_{t}\subset G_{x}$ $(x=2,\ldots,5)$, the intersection of set $G_{x-1}$ and PI set $R_{t}$ is exhibited in table \ref{33}. This shows that the condition iii) is true.
\begin{table}[tbp]
\centering
\caption{The intersection of set $G_{x-1}$ and PI set $R_{t}$ about subset $U_{t}\subset G_{x}$ $(x=2,\ldots,5)$.}\label{33}
\begin{tabular}{cc|cc}
\hline
\hline
~~~Subset~~~ & ~~~~~Intersection~~~~~ & ~~~Subset~~~ & ~~~~~Intersection~~~~~   \\ \hline
 $U_{11}\subset G_{2}$ & $U_{12}=G_{1}\cap R_{11}$ & $U_{44}\subset G_{3}$ & $U_{11}=G_{2}\cap R_{44}$        \\
 $U_{13}\subset G_{2}$ & $U_{31}=G_{1}\cap R_{13}$ & $U_{52}\subset G_{3}$ & $U_{11}=G_{2}\cap R_{52}$        \\
 $U_{42}\subset G_{2}$ & $U_{61}=G_{1}\cap R_{42}$ & $U_{91}\subset G_{3}$ & $U_{51}=G_{2}\cap R_{91}$        \\
 $U_{51}\subset G_{2}$ & $U_{62}=G_{1}\cap R_{51}$ & $U_{24}\subset G_{4}$ & $U_{52}=G_{3}\cap R_{24}$        \\
 $U_{54}\subset G_{2}$ & $U_{21}=G_{1}\cap R_{54}$ & $U_{43}\subset G_{4}$ & $U_{22}=G_{3}\cap R_{43}$ \\
 $U_{63}\subset G_{2}$ & $U_{12}=G_{1}\cap R_{63}$ & $U_{53}\subset G_{4}$ & $U_{32}=G_{3}\cap R_{53}$\\
 $U_{14}\subset G_{3}$ & $U_{42}=G_{2}\cap R_{14}$ & $U_{71}\subset G_{4}$ & $U_{41}=G_{3}\cap R_{71}$\\
 $U_{22}\subset G_{3}$ & $U_{13}=G_{2}\cap R_{22}$ & $U_{72}\subset G_{4}$ & $U_{14}=G_{3}\cap R_{72}$\\
 $U_{23}\subset G_{3}$ & $U_{11}=G_{2}\cap R_{23}$ & $U_{92}\subset G_{5}$ & $U_{24}=G_{4}\cap R_{92}$\\
 $U_{41}\subset G_{3}$ & $U_{11}=G_{2}\cap R_{41}$ &\\
\hline
\end{tabular}
\end{table}

We find the tree sequence of projection sets $U_{12}^{(BCD)}\rightarrow U_{61}^{(BCD)}(\rightarrow U_{42}^{(BCD)}\rightarrow U_{14}^{(BCD)})\rightarrow U_{31}^{(BCD)}\rightarrow U_{32}^{(BCD)}\rightarrow U_{51}^{(BCD)}(\rightarrow U_{62}^{(BCD)}\rightarrow U_{34}^{(BCD)})\rightarrow U_{21}^{(BCD)}\rightarrow U_{22}^{(BCD)}\rightarrow U_{41}^{(BCD)}(\rightarrow U_{11}^{(BCD)})\rightarrow U_{52}^{(BCD)}\rightarrow U_{24}^{(BCD)}$, where the subsequence in parentheses is a branch of the previous adjacent set. In this sequence, the intersection of the sets on both sides of the arrow is nonempty and the union of all these sets is the computation basis $\mathcal{B}^{BCD}$. This means that the family of projection sets $\{U_{ij}^{(BCD)}\}_{ij}$ is connected. The condition iv) is proven.

Therefore, one can only perform a trivial orthogonality-preserving POVM on the $BCD$ party. Combining Lemma 1 with the symmetry of (\ref{24}) ensures that the OPS (\ref{24}) is of the strongest quantum nonlocality.

\section{The proof of theorem 8}\label{H}
Suppose that the whole quantum system is shared among Alice, Bob, and Charlie.
 By taking advantage of entangled resource $|\phi^{+}(d_{C})\rangle$, Charlie first teleports the state in his subsystem $C$ to Bob. Let the subindex $\widetilde{B}$ represent the joint part of  $B$ and $C$. Whereafter, to locally discriminate the states in ({\ref{21}}), the EPR state $|\phi^{+}(2)\rangle_{ab}$ is shared by Alice and Bob. The initial state is
\begin{equation}\label{41}
\begin{aligned}
|\psi\rangle_{A\widetilde{B}}\otimes|\phi^{+}(2)\rangle_{ab},
\end{aligned}
\end{equation}
where $|\psi\rangle_{A\widetilde{B}}$ is one of the states from the set ({\ref{21}}), $a$ and $b$ are ancillary systems of Alice and Bob, respectively. Because each subset $H_{r}$ $(r\in Q)$ is LOCC distinguishable, one only needs to locally distinguish these subsets. Now the discrimination protocol proceeds as follows.

\begin{figure}[h]
\centering
\includegraphics[width=0.48\textwidth]{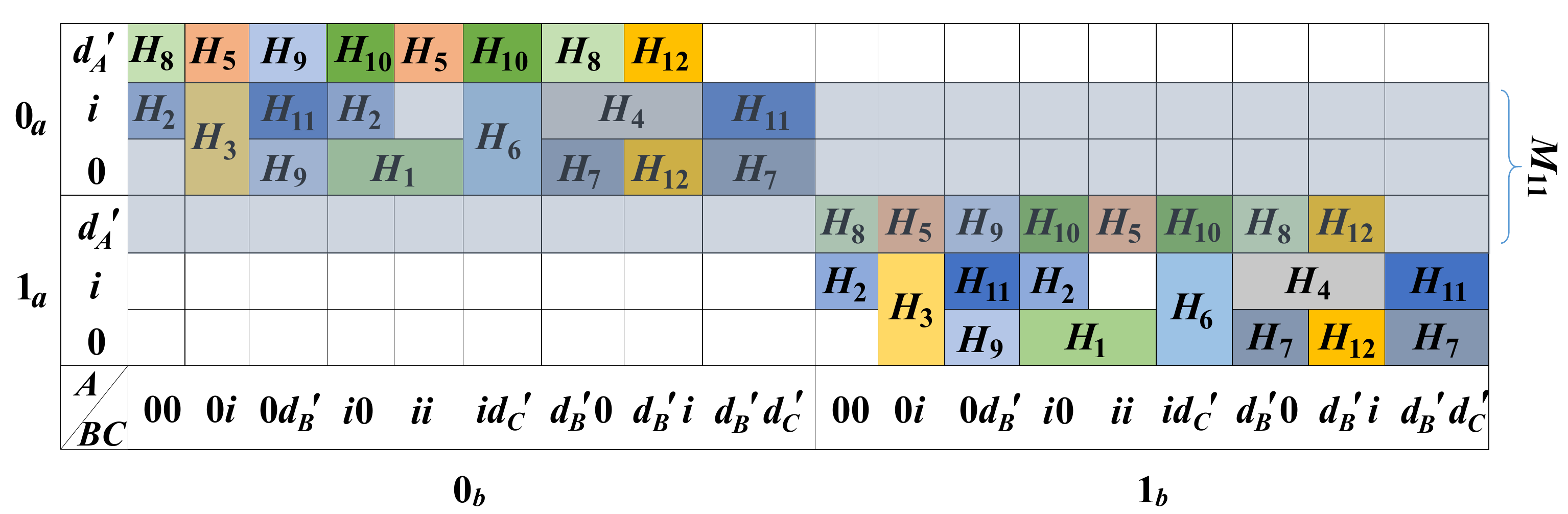}
\caption{While Alice and Bob share the EPR state $|\phi^{+}(2)\rangle_{ab}$, the initial state given by Eq. (\ref{41}) can be expressed by the corresponding $2d_{A}\times 2d_{B}d_{C}$ grid. Area covered with light gray represents the measurement effect $M_{11}$ in step 1. \label{34}}
\end{figure}

$Step~1.$ Alice performs the measurement
\begin{equation*}
\begin{aligned}
\mathcal{M}_{1}\equiv\{&M_{11}:=P[(|0\rangle,\ldots,|d_{A}'-1\rangle)_{A};|0\rangle_{a}]\\
                       &\quad\quad\quad+P[|d_{A}'\rangle_{A};|1\rangle_{a}],~\\
                       &M_{12}:=I-M_{11}\},
\end{aligned}
\end{equation*}
where $P[(|0\rangle,\ldots,|d_{A}'-1\rangle)_{A};|0\rangle_{a}]:=(|0\rangle\langle 0|+\cdots+|d_{A}'-1\rangle\langle d_{A}'-1|)_A\otimes(|0\rangle\langle 0|)_{a}$, this definition is applicable for all the protocols.
Suppose the outcome corresponding to $M_{11}$ clicks (see Fig. \ref{34}), then the resulting postmeasurement states are
\begin{equation*}
\begin{aligned}
&H_{1} \rightarrow \{|0\rangle_{A}|\xi_{i}\circ \eta_{j}\rangle_{\widetilde{B}}|00\rangle_{ab}\}, \\
&H_{2} \rightarrow \{|\xi_{i}\rangle_{A}|\eta_{j}\circ 0\rangle_{\widetilde{B}}|00\rangle_{ab}\}, \\
&H_{3} \rightarrow \{|\eta_{j}\rangle_{A}|0\circ \xi_{i}\rangle_{\widetilde{B}}|00\rangle_{ab}\}, \\
&H_{4} \rightarrow \{|\xi_{i}\rangle_{A}|d_{B}'\circ \eta_{j}\rangle_{\widetilde{B}}|00\rangle_{ab}\}, \\
&H_{5} \rightarrow \{|d_{A}'\rangle_{A}|\eta_{j}\circ \xi_{i}\rangle_{\widetilde{B}}|11\rangle_{ab}\}, \\
&H_{6} \rightarrow \{|\eta_{j}\rangle_{A}|\xi_{i}\circ d_{C}'\rangle_{\widetilde{B}}|00\rangle_{ab}\}, \\
&H_{7} \rightarrow \{|0\rangle_{A}|d_{B}'\circ (0\pm d_{C}')\rangle_{\widetilde{B}}|00\rangle_{ab}\}, \\
&H_{8} \rightarrow \{|d_{A}'\rangle_{A}|(0\pm d_{B}')\circ 0\rangle_{\widetilde{B}}|11\rangle_{ab}\}, \\
&H_{9} \rightarrow \{(|0\rangle_{A}|00\rangle_{ab}\pm |d_{A}'\rangle_{A}|11\rangle_{ab})|0\circ d_{C}'\rangle_{\widetilde{B}}\}, \\
&H_{10} \rightarrow \{|d_{A}'\rangle_{A}|\xi_{i}\circ (0\pm d_{C}')\rangle_{\widetilde{B}}|11\rangle_{ab}\}, \\
&H_{11} \rightarrow \{|\xi_{i}\rangle_{A}|(0\pm d_{B}')\circ d_{C}'\rangle_{\widetilde{B}}|00\rangle_{ab}\}, \\
&H_{12} \rightarrow \{(|0\rangle_{A}|00\rangle_{ab}\pm |d_{A}'\rangle_{A}|11\rangle_{ab})|d_{B}'\circ \xi_{i}\rangle_{\widetilde{B}}\}.
\end{aligned}
\end{equation*}
Henceforth, symbol `$\circ$' represents the union of the parties. For example, $|\psi_{1}\circ \psi_{2}\rangle_{\widetilde{B}}=|\psi_{1}\rangle_{B}|\psi_{2}\rangle_{C}$ for any two quantum states $|\psi_{1}\rangle_{B}$ and $|\psi_{2}\rangle_{C}$. Specially, let $|(0,\ldots,d_{B}-1)\circ (0,\ldots,d_{C}-1)\rangle_{\widetilde{B}}$ express the set $\{|ij\rangle_{\widetilde{B}}~ |~i=0,1,\cdots,d_B-1; j=0,1,\cdots,d_C-1 \}$ denoted by $(|00\rangle,\ldots,|0(d_{C}-1)\rangle,|10\rangle,\ldots,|(d_{B}-1)(d_{C}-1)\rangle)_{\widetilde{B}}$.

\begin{figure}[h]
\centering
\includegraphics[width=0.48\textwidth]{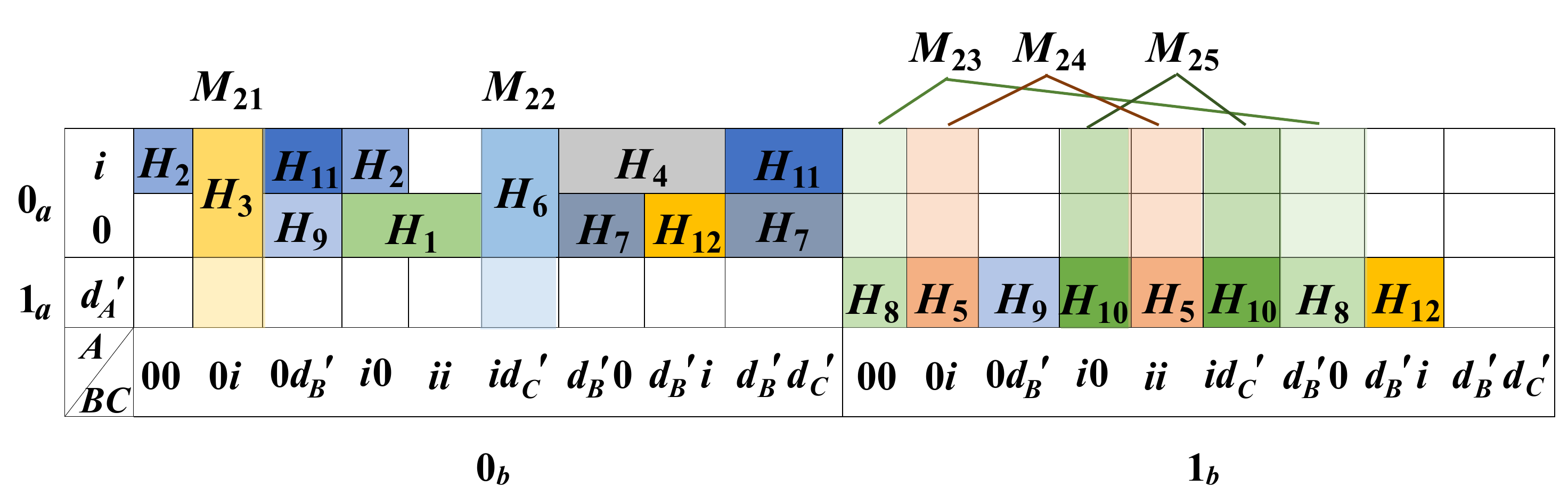}
\caption{The $d_{A}\times 2d_{B}d_{C}$ grid is the states after clicking $M_{11}$. Areas covered by different light colors denote the different measurement effect. \label{35}}
\end{figure}

$Step~2.$ Bob performs the measurement
\begin{equation*}
\begin{aligned}
\mathcal{M}_{2}\equiv\{&M_{21}:=P[|0\circ (1,\ldots,d_{C}'-1)\rangle_{\widetilde{B}};|0\rangle_{b}],~\\
&M_{22}:=P[|(1,\ldots,d_{B}'-1)\circ d_{C}'\rangle_{\widetilde{B}};|0\rangle_{b}],~\\
&M_{23}:=P[|(0,d_{B}')\circ 0\rangle_{\widetilde{B}};|1\rangle_{b}],~\\
\end{aligned}
\end{equation*}
\begin{equation*}
\begin{aligned}
&M_{24}:=P[|(0,\ldots,d_{B}'-1)\circ (1,\ldots,d_{C}'-1)\rangle_{\widetilde{B}};\\
&\quad\quad\quad\quad~~|1\rangle_{b}],~\\
&M_{25}:=P[|(1,\ldots,d_{B}'-1)\circ (0,d_{C}')\rangle_{\widetilde{B}};|1\rangle_{b}],~\\
&M_{26}:=I-M_{1}-M_{2}-M_{3}-M_{4}-M_{5}\}.
\end{aligned}
\end{equation*}
This step is shown in Fig. \ref{35}. If the corresponding operations $M_{21}$, $M_{22}$, $M_{23}$, $M_{24}$ and $M_{25}$ click, we can distinguish the subsets $H_{3}$, $H_{6}$, $H_{8}$, $H_{5}$ and $H_{10}$, respectively. If $M_{26}$ clicks, the given state is belonging to one of the remaining seven subsets $\{H_{1},H_{2},H_{4},H_{7},H_{9},H_{11},H_{12}\}$. At this point, we move on to the next step.

\begin{figure}[h]
\centering
\includegraphics[width=0.44\textwidth]{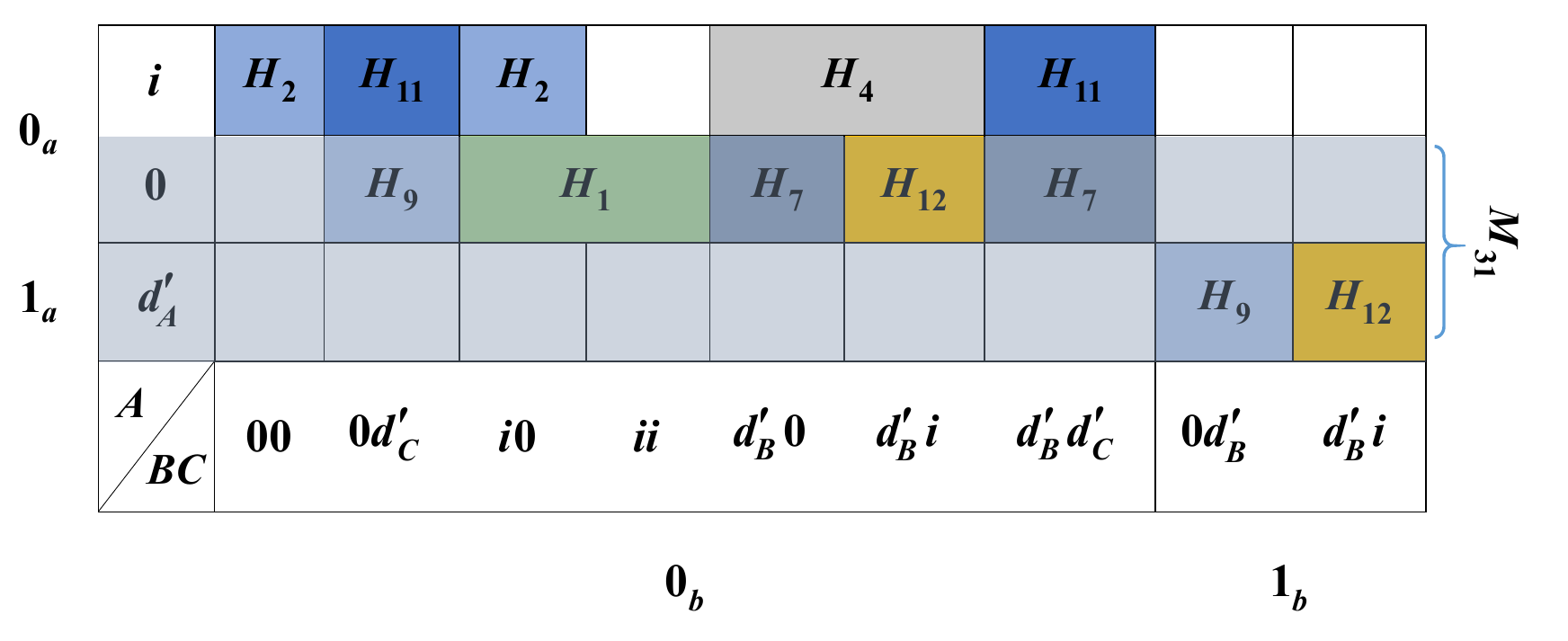}
\caption{The remaining states after Bob performs the measurement. Area covered by light gray is the measurement effect $M_{31}$. \label{38}}
\end{figure}

$Step~3.$ Alice performs the measurement
\begin{equation*}
\begin{aligned}
\mathcal{M}_{3}\equiv\{&M_{31}:=P[|0\rangle_{A};|0\rangle_{a}]+P[|d_{A}'\rangle_{A};|1\rangle_{a}],~\\
                       &M_{32}:=I-M_{31}\}.
\end{aligned}
\end{equation*}
Fig. \ref{38} shows the intuitive situation. If $M_{31}$ clicks, we can determine the four subsets $\{H_{1}, H_{7}, H_{9}, H_{12}\}$. Otherwise, the subset is one of the remaining three $\{H_{2}, H_{4}, H_{11}\}$. Moreover, they are all perfectly LOCC distinguishable.

In addition, if $M_{12}$ clicks in step 1, we can find a similar protocol where these states can be perfectly LOCC distinguished.

\section{The proof of theorem 9}\label{I}
Naturally, we only need to locally distinguish these subsets. To this end, let Alice and Bob share an EPR state $|\phi^{+}(2)\rangle_{a_{1}b_{1}}$, meanwhile Alice and Charlie share the EPR state $|\phi^{+}(2)\rangle_{a_{2}c_{1}}$. Therefore, the initial state is
\begin{equation}\label{42}
\begin{aligned}
|\psi\rangle_{ABC}\otimes|\phi^{+}(2)\rangle_{a_{1}b_{1}}\otimes|\phi^{+}(2)\rangle_{a_{2}c_{1}},
\end{aligned}
\end{equation}
where the state $|\psi\rangle_{ABC}$ is one of the states from the set $\cup_{r=1}^{12}H_{r}$ ({\ref{21}}), $a_{1}$ and $a_{2}$ are ancillary systems of Alice, $b_{1}$ and $c_{1}$ are ancillary systems of Bob and Charlie, respectively. The specific process is as follows.

\begin{figure}[h]
\centering
\includegraphics[width=0.45\textwidth]{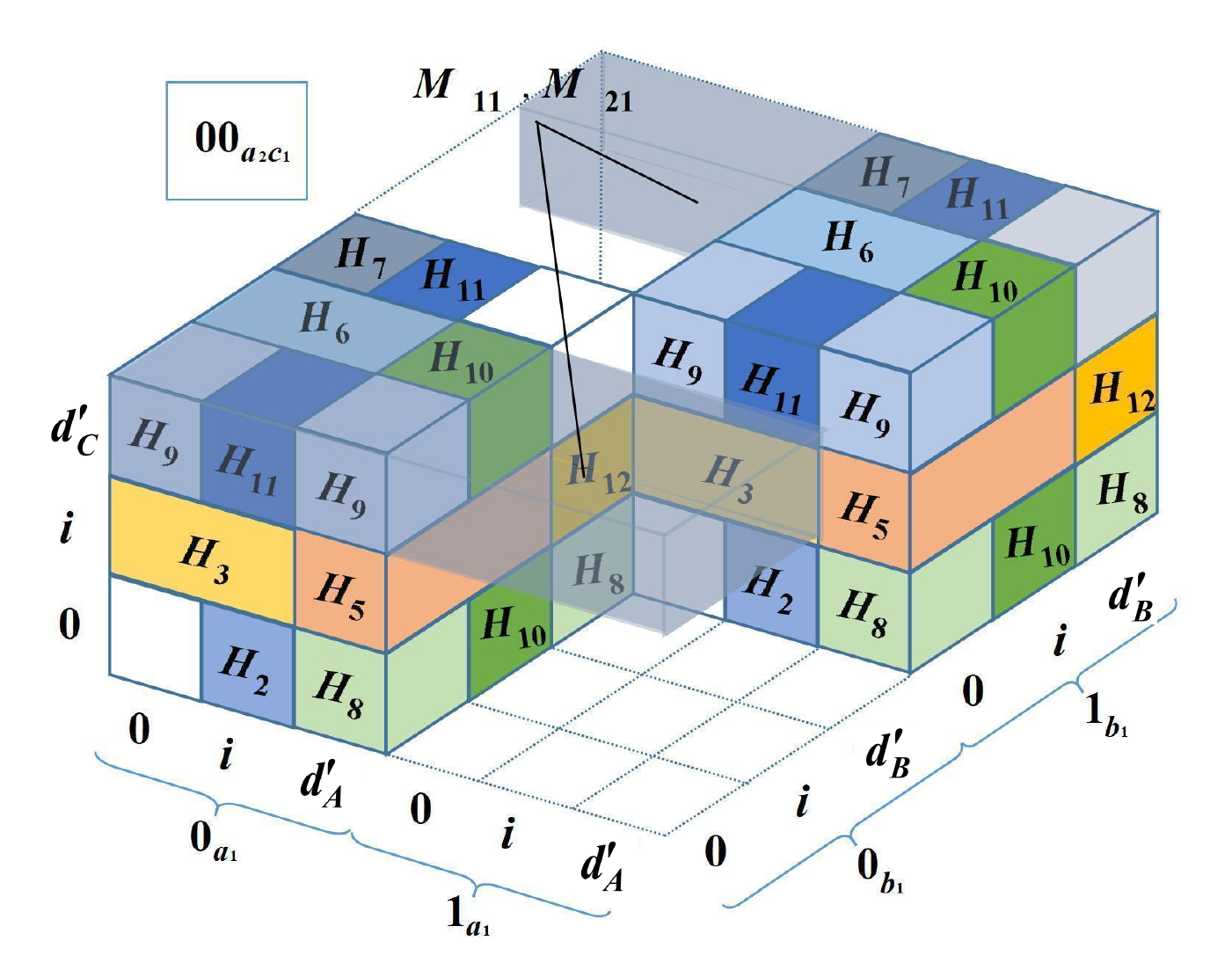}
\includegraphics[width=0.45\textwidth]{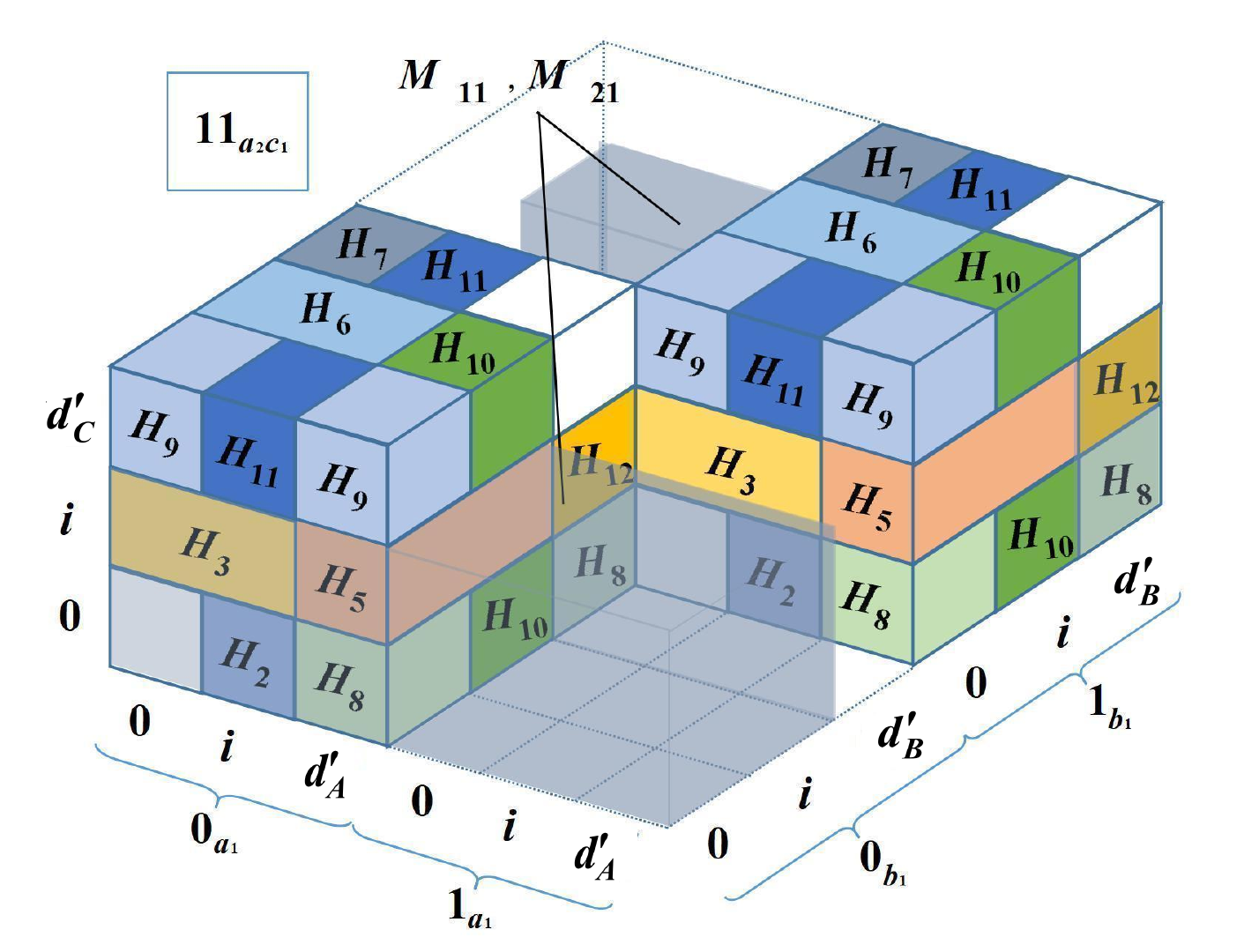}
\caption{The two $2d_{A}\times 2d_{B}\times d_{C}$ grids represent the initial states (\ref{42}) of auxiliary system as $|00\rangle_{a_{2}c_{1}}$ and $|11\rangle_{a_{2}c_{1}}$, respectively. Areas covered with light gray represent the measurement effect $M_{11}$ and $M_{21}$ in step 1. \label{36}}
\end{figure}

$Step~1.$ Bob performs the measurement
\begin{equation*}
\begin{aligned}
\mathcal{M}_{1}\equiv\{&M_{11}:=P[(|0\rangle,\ldots,|d_{B}'-1\rangle)_{B};|0\rangle_{b_{1}}]\\
                       &\quad\quad\quad+P[|d_{B}'\rangle_{B};|1\rangle_{b_{1}}],~\\
                       &M_{12}:=I-M_{11}\},
\end{aligned}
\end{equation*}
and Charlie performs the measurement
\begin{equation*}
\begin{aligned}
\mathcal{M}_{2}\equiv\{&M_{21}:=P[(|0\rangle,\ldots,|d_{C}'-1\rangle)_{C};|1\rangle_{c_{1}}]\\
                       &\quad\quad\quad+P[|d_{C}'\rangle_{C};|0\rangle_{c_{1}}],~\\
                       &M_{22}:=I-M_{21}\}.
\end{aligned}
\end{equation*}
Suppose $M_{11}$ and $M_{21}$ click (refer to Fig. \ref{36}), the resulting postmeasurement states are
\begin{equation*}
\begin{aligned}
&H_{1} \rightarrow \{|0\rangle_{A}|\xi_{i}\rangle_{B}|\eta_{j}\rangle_{C}|00\rangle_{a_{1}b_{1}}|11\rangle_{a_{2}c_{1}}\},\\
&H_{2} \rightarrow \{|\xi_{i}\rangle_{A}|\eta_{j}\rangle_{B}|0\rangle_{C}|00\rangle_{a_{1}b_{1}}|11\rangle_{a_{2}c_{1}}\},\\
&H_{3} \rightarrow \{|\eta_{j}\rangle_{A}|0\rangle_{B}|\xi_{i}\rangle_{C}|00\rangle_{a_{1}b_{1}}|11\rangle_{a_{2}c_{1}}\},\\
&H_{4} \rightarrow \{|\xi_{i}\rangle_{A}|d_{B}'\rangle_{B}|\eta_{j}\rangle_{C}|11\rangle_{a_{1}b_{1}}|11\rangle_{a_{2}c_{1}}\},\\
&H_{5} \rightarrow \{|d_{A}'\rangle_{A}|\eta_{j}\rangle_{B}|\xi_{i}\rangle_{C}|00\rangle_{a_{1}b_{1}}|11\rangle_{a_{2}c_{1}}\},\\
&H_{6} \rightarrow \{|\eta_{j}\rangle_{A}|\xi_{i}\rangle_{B}|d_{C}'\rangle_{C}|00\rangle_{a_{1}b_{1}}|00\rangle_{a_{2}c_{1}}\},\\
&H_{7} \rightarrow \{|0\rangle_{A}|d_{B}'\rangle_{B}|11\rangle_{a_{1}b_{1}}(|0\rangle_{C}|11\rangle_{a_{2}c_{1}}\pm \\ &\quad\quad\quad~|d_{C}'\rangle_{C}|00\rangle_{a_{2}c_{1}})\},\\
&H_{8} \rightarrow \{|d_{A}'\rangle_{A}(|0\rangle_{B}|00\rangle_{a_{1}b_{1}}\pm |d_{B}'\rangle_{B}|11\rangle_{a_{1}b_{1}})\\
&\quad\quad\quad~|0\rangle_{C}|11\rangle_{a_{2}c_{1}}\},\\
&H_{9} \rightarrow \{|0\pm d_{A}'\rangle_{A}|0\rangle_{B}|d_{C}'\rangle_{C}|00\rangle_{a_{1}b_{1}}|00\rangle_{a_{2}c_{1}}\},\\
&H_{10} \rightarrow \{|d_{A}'\rangle_{A}|\xi_{i}\rangle_{B}|00\rangle_{a_{1}b_{1}}(|0\rangle_{C}|11\rangle_{a_{2}c_{1}}\pm \\ &\quad\quad\quad~~|d_{C}'\rangle_{C}|00\rangle_{a_{2}c_{1}})\},\\
&H_{11} \rightarrow \{|\xi_{i}\rangle_{A}(|0\rangle_{B}|00\rangle_{a_{1}b_{1}}\pm |d_{B}'\rangle_{B}|11\rangle_{a_{1}b_{1}})\\
&\quad\quad\quad~~|d_{C}'\rangle_{C}|00\rangle_{a_{2}c_{1}}\},\\
&H_{12} \rightarrow \{|0\pm d_{A}'\rangle_{A}|d_{B}'\rangle_{B}|\xi_{i}\rangle_{C}|11\rangle_{a_{1}b_{1}}|11\rangle_{a_{2}c_{1}}\}.
\end{aligned}
\end{equation*}

\begin{figure}[h]
\centering
\includegraphics[width=0.38\textwidth]{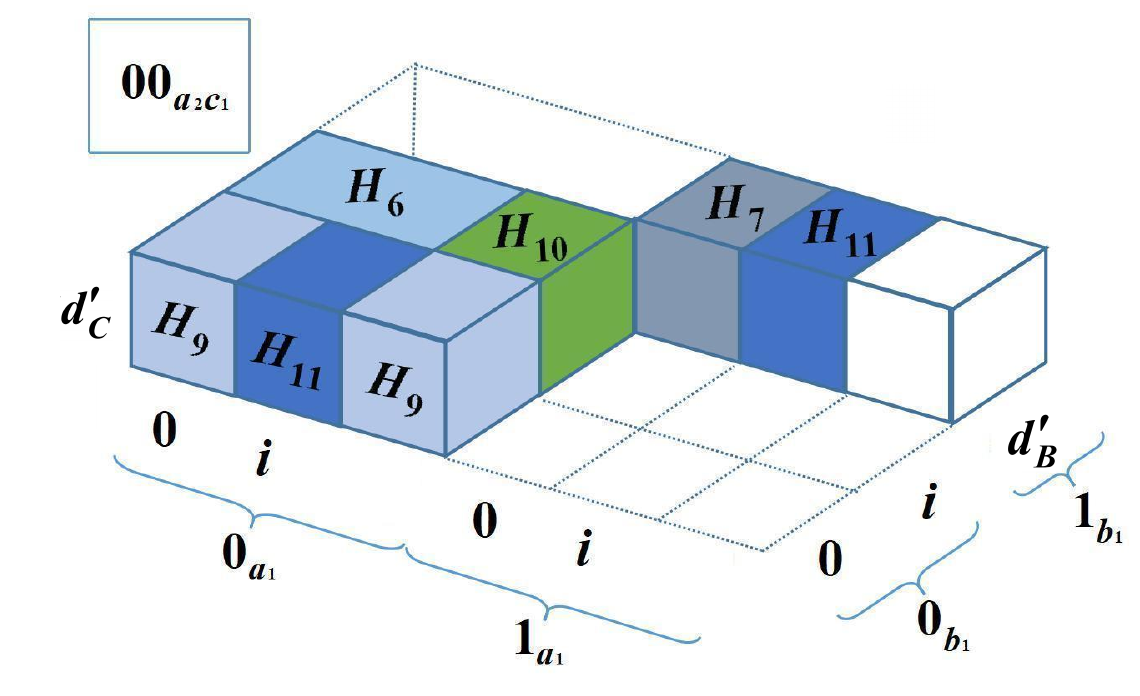}
\includegraphics[width=0.38\textwidth]{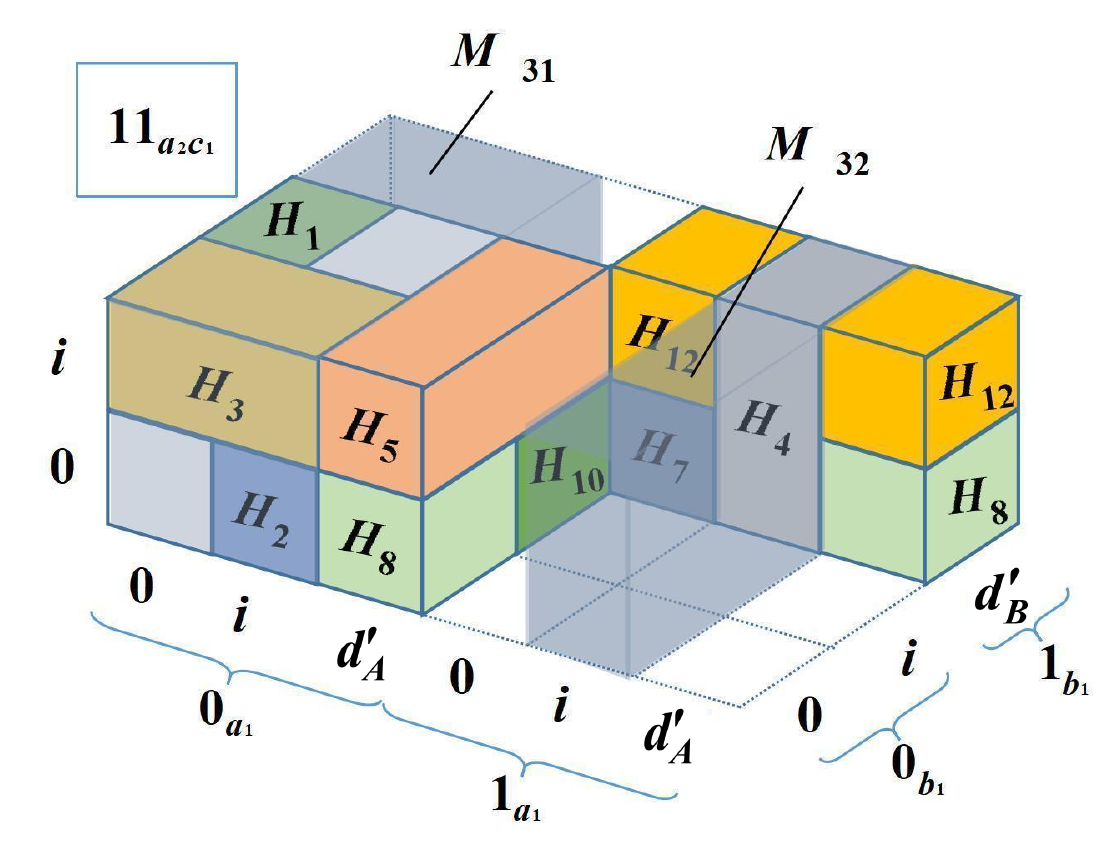}
\caption{The states after clicking $M_{11}$ and $M_{21}$. The two areas covered with light gray express the measurement effect $M_{31}$ and $M_{32}$, respectively. \label{37}}
\end{figure}

$Step~2.$ Alice performs the measurement
\begin{equation*}
\begin{aligned}
\mathcal{M}_{3}\equiv\{&M_{31}:=P[(|0\rangle,\ldots,|d_{A}'-1\rangle)_{A};|0\rangle_{a_{1}};|1\rangle_{a_{2}}],~\\
                       &M_{32}:=P[(|1\rangle,\ldots,|d_{A}'-1\rangle)_{A};|1\rangle_{a_{1}};|1\rangle_{a_{2}}],~\\
                       &M_{33}:=I-M_{31}-M_{32}\}.
\end{aligned}
\end{equation*}
This process is described in Fig. \ref{37}. If $M_{31}$ clicks, the given subset is one of $\{H_{1},H_{2},H_{3}\}$, which contains $e-3f+6$ quantum states in total. Here $e=d_{A}d_{B}+d_{A}d_{C}+d_{B}d_{C}$ and $f=d_{A}+d_{B}+d_{C}$. It is obvious that these three subsets can not be perfectly distinguished by LOCC. Let Alice and Bob share the maximally entangled state $|\phi^{+}(2)\rangle_{a_{3}b_{2}}$. Moreover, Bob performs the measurement $\mathcal{M}_{3}'\equiv\{M_{31}':=P[|0\rangle_{B};I_{b_{1}};|0\rangle_{b_{2}}]+P[(|1\rangle,\ldots,|d_{B}'\rangle)_{B};I_{b_{1}};|1\rangle_{b_{2}}],~M_{32}':=I-M_{31}'\}$. When $M_{31}'$ clicks, Alice performs the measurement $\mathcal{M}_{3}''\equiv\{M_{31}'':=P[|0\rangle_{A};I_{a_{1}};I_{a_{2}};|1\rangle_{a_{3}}],~M_{32}'':=I-M_{31}''\}$. The results corresponding to operators $M_{31}''$ and $M_{32}''$ are $H_{1}$ and $\{H_{2},H_{3}\}$, respectively. The collection $\{H_{2},H_{3}\}$ is LOCC distinguishable. Similarly, when $M_{32}'$ clicks, the task of local discrimination can also be accomplished.  The average entanglement consumed in this process is $\frac{e-3f+6}{2e-4f+6}$ maximally entangled state $|\phi^{+}(2)\rangle_{a_{3}b_{2}}$ \cite{Rout}, because the size of the set ({\ref{21}}) is $2e-4f+6$.

If $M_{32}$ clicks, the subset is $H_{4}$. Otherwise, the subset is one of the remaining eight.

\begin{figure}[h]
\centering
\includegraphics[width=0.34\textwidth]{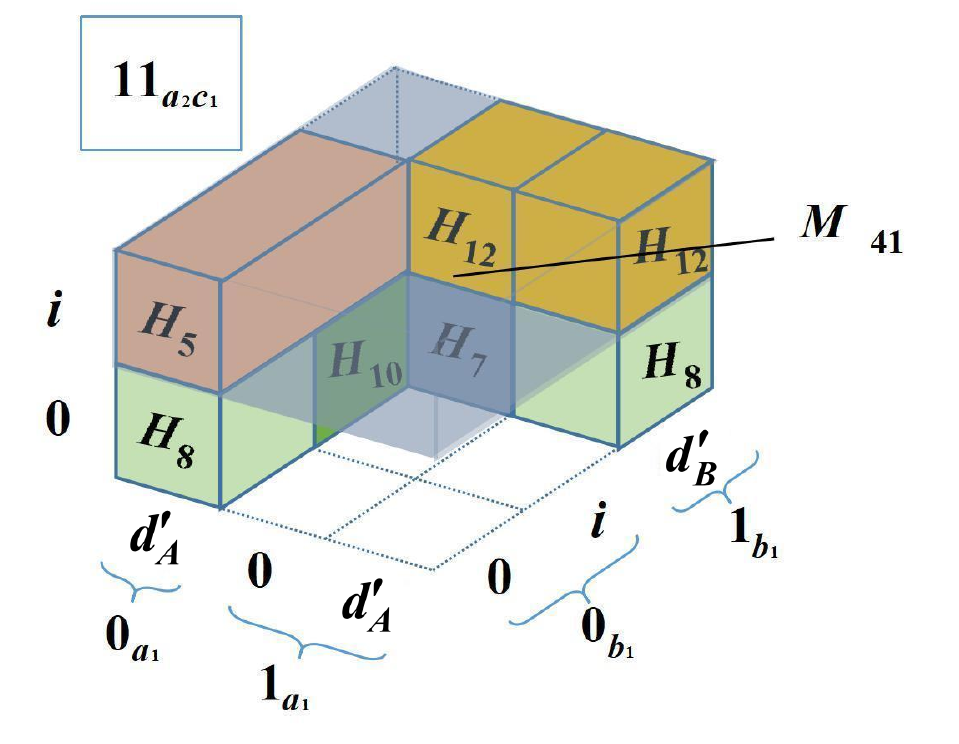}
\caption{The states with auxiliary system $|11\rangle_{a_{2}c_{1}}$ after clicking $M_{33}$. The area covered with light gray represents the measurement effect $M_{41}$. \label{39}}
\end{figure}

$Step~3.$ Charlie performs the measurement
\begin{equation*}
\begin{aligned}
\mathcal{M}_{4}\equiv\{&M_{41}:=P[(|1\rangle,\ldots,|d_{C}'-1\rangle)_{C};|1\rangle_{c_{1}}],~\\
                       &M_{42}:=I-M_{41}\}.
\end{aligned}
\end{equation*}
Refer to Fig. \ref{39}, if $M_{41}$ clicks, the given subset is one of $\{H_{5},H_{12}\}$. Obviously, it is locally distinguishable.

$Step~4.$ Alice performs the measurement
\begin{equation*}
\begin{aligned}
\mathcal{M}_{5}\equiv\{M_{51}:=P[|0\rangle_{A};|1\rangle_{a_{1}};I_{a_{2}}],~M_{52}:=I-M_{51}\}.
\end{aligned}
\end{equation*}
If $M_{51}$ clicks, the subset is $H_{7}$. Otherwise, the subset is one of $\{H_{6},H_{8},H_{9},H_{10},H_{11}\}$.

$Step~5.$ Bob performs the measurement
\begin{equation*}
\begin{aligned}
\mathcal{M}_{6}\equiv\{&M_{61}:=P[|0\rangle_{B};|0\rangle_{b_{1}}]+P[|d_{B}'\rangle_{B};|1\rangle_{b_{1}}],~\\
                       &M_{62}:=I-M_{61}\}.
\end{aligned}
\end{equation*}
The results corresponding to operators $M_{61}$ and $M_{62}$ are $\{H_{8},H_{9},H_{11}\}$ and $\{H_{6},H_{10}\}$, respectively. They are all locally distinguishable.

In summary, we consume a total of $1+\frac{e-3f+6}{2e-4f+6}$ EPR states between Alice and Bob and one EPR state between Alice and Charlie for this distinguishing task. If in the step 1 other operators click, we can find similar protocols to distinguish these subsets perfectly by LOCC.

\section{The proof of theorem 10}\label{J}

Suppose that the whole quantum system is shared among Alice, Bob, and Charlie. Since $d_{C}\leq d_{B}$, the subsystem $C$ is teleported to Bob by using the entanglement resource $|\phi^{+}(d_{C})\rangle_{BC}$, and the new union subsystem is represented by $\widetilde{B}$. To locally discriminate the states, Alice and Bob should share a maximally entangled state $|\phi^{+}(2)\rangle_{ab}$. The discrimination protocol proceeds as follows.

$Step~1.$ Alice performs the measurement
\begin{equation*}
\begin{aligned}
\mathcal{M}_{1}\equiv\{&M_{11}:=P[(|0\rangle,|1\rangle)_{A};|0\rangle_{a}]+P[(|2\rangle,\ldots,|d_{A}'\rangle)_{A}\\
&\quad\quad\quad~;|1\rangle_{a}],~\\
                       &M_{12}:=I-M_{11}\}.
\end{aligned}
\end{equation*}
Suppose $M_{11}$ clicks, then the resulting postmeasurement states are
\begin{equation*}
\begin{aligned}
&H_{11}\rightarrow \{|0\rangle_{A}|1\circ \alpha_{3}^{l}\rangle_{\widetilde{B}}|00\rangle_{ab}\}, \\
&H_{12}\rightarrow \{|1\rangle_{A}|\alpha_{3}^{l}\circ 0\rangle_{\widetilde{B}}|00\rangle_{ab}\}, \\
&H_{13}\rightarrow \{|\alpha_{3}^{l}\rangle_{A}|0\circ 1\rangle_{\widetilde{B}}|11\rangle_{ab}\}, \\
&H_{21}\rightarrow \{|0\rangle_{A}|\alpha^{i}\circ \alpha_{1}^{k}\rangle_{\widetilde{B}}|00\rangle_{ab}\}, \\
&H_{22}\rightarrow \{|\alpha^{i}\rangle_{A}|\alpha_{1}^{k}\circ 0\rangle_{\widetilde{B}}|11\rangle_{ab}\}, \\
&H_{23}\rightarrow \{(|1\rangle_{A}|00\rangle_{ab}+|\alpha_{1}^{k,2}\rangle_{A}|11\rangle_{ab})|0\circ \alpha^{i}\rangle_{\widetilde{B}}\}, \\
&H_{31}\rightarrow \{|0\rangle_{A}|d_{B}'\circ \alpha_{0}^{j}\rangle_{\widetilde{B}}|00\rangle_{ab}\}, \\
&H_{32}\rightarrow \{|d_{A}'\rangle_{A}|\alpha_{0}^{j}\circ 0\rangle_{\widetilde{B}}|11\rangle_{ab}\}, \\
&H_{33}\rightarrow \{(|0\rangle_{A}|00\rangle_{ab}+|\alpha_{0}^{j,2}\rangle_{A}|11\rangle_{ab})|0\circ d_{C}'\rangle_{\widetilde{B}}\}, \\
&H_{41}\rightarrow \{|1\rangle_{A}|0\circ (0\pm 1)\rangle_{\widetilde{B}}|00\rangle_{ab}\}, \\
&H_{42}\rightarrow \{|0\rangle_{A}|(0\pm 1)\circ 1\rangle_{\widetilde{B}}|00\rangle_{ab}\}, \\
&H_{43}\rightarrow \{|0\pm 1\rangle_{A}|1\circ 0\rangle_{\widetilde{B}}|00\rangle_{ab}\}, \\
&H_{51}\rightarrow \{|1\rangle_{A}|d_{B}'\circ \alpha_{3}^{l}\rangle_{\widetilde{B}}|00\rangle_{ab}\}, \\
&H_{52}\rightarrow \{|d_{A}'\rangle_{A}|\alpha_{3}^{l}\circ 1\rangle_{\widetilde{B}}|11\rangle_{ab}\}, \\
&H_{53}\rightarrow \{|\alpha_{3}^{l}\rangle_{A}|1\circ d_{C}'\rangle_{\widetilde{B}}|11\rangle_{ab}\}, \\
&H_{61}\rightarrow \{|\alpha^{i}\rangle_{A}|d_{B}'\circ \alpha_{1}^{k}\rangle_{\widetilde{B}}|11\rangle_{ab}\}, \\
&H_{62}\rightarrow \{|d_{A}'\rangle_{A}|\alpha_{1}^{k}\circ \alpha^{i}\rangle_{\widetilde{B}}|11\rangle_{ab}\}, \\
&H_{63}\rightarrow \{(|1\rangle_{A}|00\rangle_{ab}+|\alpha_{1}^{k,2}\rangle_{A}|11\rangle_{ab})|\alpha^{i}\circ d_{C}'\rangle_{\widetilde{B}}\}, \\
&H_{71}\rightarrow \{|d_{A}'\rangle_{A}|0\circ \alpha_{3}^{l}\rangle_{\widetilde{B}}|11\rangle_{ab}\}, \\
&H_{72}\rightarrow \{|0\rangle_{A}|\alpha_{3}^{l}\circ d_{C}'\rangle_{\widetilde{B}}|00\rangle_{ab}\}, \\
&H_{73}\rightarrow \{|\alpha_{3}^{l}\rangle_{A}|d_{B}'\circ 0\rangle_{\widetilde{B}}|11\rangle_{ab}\}, \\
&H_{81}\rightarrow \{|d_{A}'\rangle_{A}|1\circ (0\pm 1)\rangle_{\widetilde{B}}|11\rangle_{ab}\}, \\
&H_{82}\rightarrow \{|1\rangle_{A}|(0\pm 1)\circ d_{C}'\rangle_{\widetilde{B}}|00\rangle_{ab}\}, \\
&H_{83}\rightarrow \{|0\pm 1\rangle_{A}|d_{B}'\circ 1\rangle_{\widetilde{B}}|00\rangle_{ab}\},
\end{aligned}
\end{equation*}
where $|\alpha_{0}^{j,2}\rangle_{A}=\sum_{u=1}^{d_{A}-3}\omega_{d_{A}-2}^{ju}|u+1\rangle$ and $|\alpha_{1}^{k,2}\rangle_{A}=\sum_{u=1}^{d_{A}-3}\omega_{d_{A}-2}^{ku}|u+1\rangle$.

$Step~2.$ Bob performs the measurement
\begin{equation*}
\begin{aligned}
\mathcal{M}_{2}\equiv\{&M_{21}:=P[|(1,\ldots,d_{B}'-1)\circ (2,\ldots,d_{C}'-1)\rangle_{\widetilde{B}}\\
&\quad\quad\quad~;|1\rangle_{b}],~\\
&M_{22}:=P[(|(2,\ldots,d_{B}')\circ (0,d_{C}')\rangle,|d_{B}'\circ (2,\ldots,\\
&\quad\quad\quad~ d_{C}'-1)\rangle)_{\widetilde{B}};|0\rangle_{b}]+P[|(2,\ldots,d_{B}'-1)\\
&\quad\quad\quad~\circ d_{C}'\rangle_{\widetilde{B}};|1\rangle_{b}],~\\
&M_{23}:=P[|01\rangle_{\widetilde{B}};|1\rangle_{b}],~\\
&M_{24}:=P[(|(0,\ldots,d_{B}'-1)\circ 0\rangle,|11\rangle)_{\widetilde{B}};|1\rangle_{b}],~\\
\end{aligned}
\end{equation*}
\begin{equation*}
\begin{aligned}
&M_{25}:=P[|1d_{C}'\rangle_{\widetilde{B}};|1\rangle_{b}],~\\
&M_{26}:=P[(|(2,\ldots,d_{B}')\circ 1\rangle,|d_{B}'2\rangle)_{\widetilde{B}};|1\rangle_{b}],~\\
&M_{27}:=P[|d_{B}'0\rangle_{\widetilde{B}};|1\rangle_{b}],~\\
&M_{28}:=P[(|00\rangle,|01\rangle,|11\rangle)_{\widetilde{B}};|0\rangle_{b}],~\\
&M_{29}:=P[|10\rangle_{\widetilde{B}};|0\rangle_{b}],~\\
&M_{210}:=P[|d_{B}'1\rangle_{\widetilde{B}};|0\rangle_{b}],~\\
&M_{211}:=P[|(2,\ldots,d_{B}'-1)\circ (1,\ldots,d_{C}'-1)\rangle_{\widetilde{B}}\\
&\quad\quad\quad~~~;|0\rangle_{b}],~\\
&M_{212}:=I-\sum_{i=1}^{11}M_{2i}\}.
\end{aligned}
\end{equation*}
For the operator $M_{2i}$ $(i=1,\ldots,12)$, the result of postmeasurement is
\begin{equation*}
\begin{aligned}
&M_{21}\Rightarrow H_{62},&&M_{22}\Rightarrow H_{12},H_{31},H_{51},H_{72},H_{63},\\
&M_{23}\Rightarrow H_{13},&&M_{24}\Rightarrow H_{22},H_{32},H_{81},\\
&M_{25}\Rightarrow H_{53},&&M_{26}\Rightarrow H_{52},H_{61},\\
&M_{27}\Rightarrow H_{73},&&M_{28}\Rightarrow H_{41},H_{42},\\
&M_{29}\Rightarrow H_{43},&&M_{210}\Rightarrow H_{83},\\
&M_{211}\Rightarrow H_{21},&&M_{212}\Rightarrow H_{11},H_{23},H_{33},H_{71},H_{82}.
\end{aligned}
\end{equation*}
Clearly, $\{H_{52},H_{61}\}$ and $\{H_{41},H_{42}\}$ are locally distinguishable. If $M_{22}$ clicks, Alice performs the measurement $\mathcal{M}_{2}'\equiv\{M_{21}':=P[|0\rangle_{A};|0\rangle_{a}],~M_{22}':=I-M_{21}'\}$. The outcomes corresponding to the operators $M_{21}'$ and $M_{22}'$ are $\{H_{31},H_{72}\}$ and $\{H_{12},H_{51},H_{63}\}$, respectively. They are also locally distinguishable. If $M_{24}$ clicks, Alice performs the measurement $\mathcal{M}_{2}''\equiv\{M_{21}'':=P[(|2\rangle,\ldots,|d_{A}'-1\rangle)_{A};|1\rangle_{a}],~M_{22}'':=I-M_{21}''\}$. The outcomes corresponding to the operators $M_{21}''$ and $M_{22}''$ are $H_{22}$ and $\{H_{32},H_{81}\}$, respectively. Moreover, $\{H_{32},H_{81}\}$ is a LOCC distinguishable collection. If $M_{212}$ clicks, we proceed to the next step.

$Step~3.$ Alice performs the measurement
\begin{equation*}
\begin{aligned}
\mathcal{M}_{3}\equiv\{M_{31}:=P[|d_{A}'\rangle_{A};|1\rangle_{a}],~M_{32}:=I-M_{31}\}.
\end{aligned}
\end{equation*}
If $M_{31}$ clicks, the subset is $H_{71}$. If $M_{32}$ clicks, the subset is one of the remaining four.

$Step~4.$ Bob performs the measurement
\begin{equation*}
\begin{aligned}
\mathcal{M}_{4}\equiv\{&M_{41}:=P[|0\circ (2,\ldots,d_{C}'-1)\rangle_{\widetilde{B}};I_{b}],~\\
                       &M_{42}:=I-M_{41}\}.
\end{aligned}
\end{equation*}
If $M_{41}$ clicks, the subset is $H_{23}$. If $M_{42}$ clicks, the result is one of the three remaining subsets.

$Step~5.$ Alice performs the measurement
\begin{equation*}
\begin{aligned}
\mathcal{M}_{5}\equiv\{M_{51}:=P[|1\rangle_{A};|0\rangle_{a}],~M_{52}:=I-M_{51}\}.
\end{aligned}
\end{equation*}
If $M_{51}$ clicks, the subset is $H_{82}$. If $M_{52}$ clicks, the subset is one of $\{H_{33},H_{11}\}$, which is locally distinguishable.

On the other hand, when $M_{12}$ clicks in the step 1, we can  find the distinction protocol similarly.

\section{The proof of theorem 11}\label{K}
To locally distinguish the set ({\ref{23}}), let Alice and Bob share a maximally entangled state $|\phi^{+}(4)\rangle_{a_{1}b_{1}}$, while Alice and Charlie share an EPR state $|\phi^{+}(2)\rangle_{a_{2}c_{1}}$.

$Step~1.$ Bob performs the measurement
\begin{equation*}
\begin{aligned}
\mathcal{M}_{1}\equiv\{&M_{11}:=P[|0\rangle_{B};|0\rangle_{b_{1}}]+P[|1\rangle_{B};|1\rangle_{b_{1}}]\\
                       &\quad\quad\quad~+P[(|2\rangle,\ldots,|d_{B}'-1\rangle)_{B};|2\rangle_{b_{1}}]\\
                       &\quad\quad\quad~+ P[|d_{B}'\rangle_{B};|3\rangle_{b_{1}}],~\\
                       &M_{12}:=P[|0\rangle_{B};|1\rangle_{b_{1}}]+P[|1\rangle_{B};|2\rangle_{b_{1}}]\\
                       &\quad\quad\quad~+P[(|2\rangle,\ldots,|d_{B}'-1\rangle)_{B};|3\rangle_{b_{1}}]\\
                       &\quad\quad\quad~+P[|d_{B}'\rangle_{B};|0\rangle_{b_{1}}],~\\
                       &M_{13}:=P[|0\rangle_{B};|2\rangle_{b_{1}}]+P[|1\rangle_{B};|3\rangle_{b_{1}}]\\
                       &\quad\quad\quad~+P[(|2\rangle,\ldots,|d_{B}'-1\rangle)_{B};|0\rangle_{b_{1}}]\\
                       &\quad\quad\quad~+ P[|d_{B}'\rangle_{B};|1\rangle_{b_{1}}],~\\
                       &M_{14}:=I-M_{11}-M_{12}-M_{13}\}.
\end{aligned}
\end{equation*}
Charlie performs the measurement
\begin{equation*}
\begin{aligned}\mathcal{M}_{2}\equiv\{&M_{21}:=P[(|0\rangle,|1\rangle)_{C};|0\rangle_{c_{1}}]+P[(|2\rangle,\\
                                      &\quad\quad\quad\quad~~\ldots,|d_{C}'\rangle)_{C};|1\rangle_{c_{1}}],~\\
                                      &M_{22}:=I-M_{21}\}.
\end{aligned}
\end{equation*}
Suppose the outcomes corresponding to $M_{11}$ and $M_{21}$ click, the resulting postmeasurement states are
\begin{equation}\label{K1}
\begin{aligned}
&H_{11}\rightarrow\{|0\rangle_{A}|1\rangle_{B}|\alpha_{3}^{l}\rangle_{C}|11\rangle_{a_{1}b_{1}}|11\rangle_{a_{2}c_{1}}\}, \\
&H_{12}\rightarrow\{|1\rangle_{A}(|\alpha_{3}^{l,1}\rangle_{B}|22\rangle_{a_{1}b_{1}}+ |\alpha_{3}^{l,2}\rangle_{B}|33\rangle_{a_{1}b_{1}})\\
&\quad\quad\quad~~|0\rangle_{C}|00\rangle_{a_{2}c_{1}}\}, \\
&H_{13}\rightarrow\{|\alpha_{3}^{l}\rangle_{A}|0\rangle_{B}|1\rangle_{C}|00\rangle_{a_{1}b_{1}}|00\rangle_{a_{2}c_{1}}\}, \\
&H_{21}\rightarrow\{|0\rangle_{A}|\alpha^{i}\rangle_{B}|22\rangle_{a_{1}b_{1}}(|1\rangle_{C}|00\rangle_{a_{2}c_{1}}+ |\alpha_{1}^{k,2}\rangle_{C}\\
&\quad\quad\quad~~|11\rangle_{a_{2}c_{1}})\}, \\
&H_{22}\rightarrow\{|\alpha^{i}\rangle_{A}(|1\rangle_{B}|11\rangle_{a_{1}b_{1}}+ |\alpha_{1}^{k,2}\rangle_{B}|22\rangle_{a_{1}b_{1}})\\
&\quad\quad\quad~~|0\rangle_{C}|00\rangle_{a_{2}c_{1}}\}, \\
&H_{23}\rightarrow\{|\alpha_{1}^{k}\rangle_{A}|0\rangle_{B}|\alpha^{i}\rangle_{C}|00\rangle_{a_{1}b_{1}}|11\rangle_{a_{2}c_{1}}\}, \\
&H_{31}\rightarrow\{|0\rangle_{A}|d_{B}'\rangle_{B}|33\rangle_{a_{1}b_{1}}(|0\rangle_{C}|00\rangle_{a_{2}c_{1}}+ |\alpha_{0}^{j,2}\rangle_{C}\\
&\quad\quad\quad~~|11\rangle_{a_{2}c_{1}})\}, \\
&H_{32}\rightarrow\{|d_{A}'\rangle_{A}(|0\rangle_{B}|00\rangle_{a_{1}b_{1}}+ |\alpha_{0}^{j,2}\rangle_{B}|22\rangle_{a_{1}b_{1}})\\
&\quad\quad\quad~~|0\rangle_{C}|00\rangle_{a_{2}c_{1}}\}, \\
&H_{33}\rightarrow\{|\alpha_{0}^{j}\rangle_{A}|0\rangle_{B}|d_{C}'\rangle_{C}|00\rangle_{a_{1}b_{1}}|11\rangle_{a_{2}c_{1}}\}, \\
&H_{41}\rightarrow\{|1\rangle_{A}|0\rangle_{B}|0\pm 1\rangle_{C}|00\rangle_{a_{1}b_{1}}|00\rangle_{a_{2}c_{1}}\}, \\
&H_{42}\rightarrow\{|0\rangle_{A}(|0\rangle_{B}|00\rangle_{a_{1}b_{1}}\pm |1\rangle_{B}|11\rangle_{a_{1}b_{1}})|1\rangle_{C}\\
&\quad\quad\quad~~|00\rangle_{a_{2}c_{1}}\}, \\
&H_{43}\rightarrow\{|0\pm 1\rangle_{A}|1\rangle_{B}|0\rangle_{C}|11\rangle_{a_{1}b_{1}}|00\rangle_{a_{2}c_{1}}\}, \\
&H_{51}\rightarrow\{|1\rangle_{A}|d_{B}'\rangle_{B}|\alpha_{3}^{l}\rangle_{C}|33\rangle_{a_{1}b_{1}}|11\rangle_{a_{2}c_{1}}\}, \\
&H_{52}\rightarrow\{|d_{A}'\rangle_{A}(|\alpha_{3}^{l,1}\rangle_{B}|22\rangle_{a_{1}b_{1}}+ |\alpha_{3}^{l,2}\rangle_{B}|33\rangle_{a_{1}b_{1}})\\
&\quad\quad\quad~~|1\rangle_{C}|00\rangle_{a_{2}c_{1}}\}, \\
&H_{53}\rightarrow\{|\alpha_{3}^{l}\rangle_{A}|1\rangle_{B}|d_{C}'\rangle_{C}|11\rangle_{a_{1}b_{1}}|11\rangle_{a_{2}c_{1}}\}, \\
&H_{61}\rightarrow\{|\alpha^{i}\rangle_{A}|d_{B}'\rangle_{B}|33\rangle_{a_{1}b_{1}}(|1\rangle_{C}|00\rangle_{a_{2}c_{1}}+ |\alpha_{1}^{k,2}\rangle_{C}\\
&\quad\quad\quad~~|11\rangle_{a_{2}c_{1}})\}, \\
&H_{62}\rightarrow\{|d_{A}'\rangle_{A}(|1\rangle_{B}|11\rangle_{a_{1}b_{1}}+ |\alpha_{1}^{k,2}\rangle_{B}|22\rangle_{a_{1}b_{1}})\\
&\quad\quad\quad~~|\alpha^{i}\rangle_{C}|11\rangle_{a_{2}c_{1}}\}, \\
\end{aligned}
\end{equation}
\begin{equation*}
\begin{aligned}
&H_{63}\rightarrow\{|\alpha_{1}^{k}\rangle_{A}|\alpha^{i}\rangle_{B}|d_{C}'\rangle_{C}|22\rangle_{a_{1}b_{1}}|11\rangle_{a_{2}c_{1}}\}, \\
&H_{71}\rightarrow\{|d_{A}'\rangle_{A}|0\rangle_{B}|\alpha_{3}^{l}\rangle_{C}|00\rangle_{a_{1}b_{1}}|11\rangle_{a_{2}c_{1}}\}, \\
&H_{72}\rightarrow\{|0\rangle_{A}(|\alpha_{3}^{l,1}\rangle_{B}|22\rangle_{a_{1}b_{1}}+ |\alpha_{3}^{l,2}\rangle_{B}|33\rangle_{a_{1}b_{1}})\\
&\quad\quad\quad~~|d_{C}'\rangle_{C}|11\rangle_{a_{2}c_{1}}\}, \\
&H_{73}\rightarrow\{|\alpha_{3}^{l}\rangle_{A}|d_{B}'\rangle_{B}|0\rangle_{C}|33\rangle_{a_{1}b_{1}}|00\rangle_{a_{2}c_{1}}\},\\
&H_{81}\rightarrow\{|d_{A}'\rangle_{A}|1\rangle_{B}|0\pm 1\rangle_{C}|11\rangle_{a_{1}b_{1}}|00\rangle_{a_{2}c_{1}}\},\\
&H_{82}\rightarrow\{|1\rangle_{A}(|0\rangle_{B}|00\rangle_{a_{1}b_{1}}\pm |1\rangle_{B}|11\rangle_{a_{1}b_{1}})|d_{C}'\rangle_{C}\\
&\quad\quad\quad~~|11\rangle_{a_{2}c_{1}}\}, \\
&H_{83}\rightarrow\{|0\pm 1\rangle_{A}|d_{B}'\rangle_{B}|1\rangle_{C}|33\rangle_{a_{1}b_{1}}|00\rangle_{a_{2}c_{1}}\},
\end{aligned}
\end{equation*}
where $|\alpha_{0}^{j,2}\rangle_{\tau}=\sum_{u=1}^{d_{\tau}-3}\omega_{d_{\tau}-2}^{ju}|u+1\rangle$, $|\alpha_{1}^{k,2}\rangle_{\tau}=\sum_{u=1}^{d_{\tau}-3}\omega_{d_{\tau}-2}^{ku}|u+1\rangle$, $|\alpha_{3}^{l,1}\rangle_{\tau}=\sum_{u=0}^{d_{\tau}-4}\omega_{d_{\tau}-2}^{lu}|u+2\rangle$ and $|\alpha_{3}^{l,2}\rangle_{\tau}=\omega_{d_{\tau}-2}^{l(d_{\tau}-3)}|d_{\tau}-1\rangle$ for $j,k,l\in \mathcal{Z}_{d_{\tau}-2}$ and $\tau=B,C$.

$Step~2.$ Alice performs the measurement
\begin{equation*}
\begin{aligned}
\mathcal{M}_{3}\equiv\{&M_{31}:=P[|d_{A}'\rangle_{A};|0\rangle_{a_{1}};|1\rangle_{a_{2}}],~\\
                       &M_{32}:=P[|1\rangle_{A};|0\rangle_{a_{1}};|0\rangle_{a_{2}}],~\\ &M_{33}:=P[(|2\rangle,\ldots,|d_{A}'-1\rangle)_{A};(|1\rangle,|2\rangle)_{a_{1}};|0\rangle_{a_{2}}],~\\
                       &M_{34}:=P[|0\rangle_{A};|1\rangle_{a_{1}};|1\rangle_{a_{2}}],~\\
                       &M_{35}:=P[|d_{A}'\rangle_{A};|1\rangle_{a_{1}};|0\rangle_{a_{2}}],~\\
                       &M_{36}:=P[(|1\rangle,\ldots,|d_{A}'-1\rangle)_{A};|2\rangle_{a_{1}};|1\rangle_{a_{2}}],~\\
                       &M_{37}:=P[|1\rangle_{A};|3\rangle_{a_{1}};|1\rangle_{a_{2}}],~\\
                       &M_{38}:=I-\sum_{i=1}^{7}M_{3i}\}.
\end{aligned}
\end{equation*}
The result of postmeasurement, corresponding to the operator $M_{3i}$ $(i=1,\ldots,7)$ is
\begin{equation*}
\begin{aligned}
&M_{31}\Rightarrow H_{71},&&M_{32}\Rightarrow H_{41},&&M_{33}\Rightarrow H_{22},&&M_{34}\Rightarrow H_{11},\\
&M_{35}\Rightarrow H_{81},&&M_{36}\Rightarrow H_{63},&&M_{37}\Rightarrow H_{51}.
\end{aligned}
\end{equation*}
If $M_{38}$ clicks, we proceed to the next step.

$Step~3.$ Charlie performs the measurement
\begin{equation*}
\begin{aligned}
\mathcal{M}_{4}\equiv\{M_{41}:=P[|d_{C}'\rangle_{C};|1\rangle_{c_{1}}],~M_{42}:=I-M_{41}\}.
\end{aligned}
\end{equation*}
If $M_{41}$ clicks, the given subset is one of $\{H_{33},H_{53},H_{72},H_{82}\}$. It is locally distinguishable. Otherwise, we continue to the next step.

$Step~4.$ Alice performs the measurement
\begin{equation*}
\begin{aligned}
\mathcal{M}_{5}\equiv\{&M_{51}:=P[(|0\rangle,|1\rangle)_{A};(|0\rangle,|1\rangle)_{a_{1}};|0\rangle_{a_{2}}], \\
                       &M_{52}:=P[(|2\rangle,\ldots,|d_{A}'\rangle)_{A};(|1\rangle,|2\rangle)_{a_{1}};|1\rangle_{a_{2}}], \\
                       &M_{53}:=P[(|1\rangle,\ldots,|d_{A}'-1\rangle)_{A};|0\rangle_{a_{1}};|1\rangle_{a_{2}}], \\
                       &M_{54}:=P[|0\rangle_{A};|2\rangle_{a_{1}};I_{a_{2}}], \\
                       &M_{55}:=P[(|0\rangle,|1\rangle)_{A};|3\rangle_{a_{1}};|0\rangle_{a_{2}}]+P[|0\rangle_{A};\\
                       &\quad\quad\quad\quad~~|3\rangle_{a_{1}};|1\rangle_{a_{2}}]+P[|1\rangle_{A};|2\rangle_{a_{1}};|0\rangle_{a_{2}}],\\
                       &M_{56}:=I-\sum_{i=1}^{5}M_{5i}\}.
\end{aligned}
\end{equation*}
Corresponding to the operator $M_{5i}$ $(i=1,\ldots,6)$, there is the following result
\begin{equation*}
\begin{aligned}
&M_{51}\Rightarrow H_{43},H_{42},&&M_{54}\Rightarrow H_{21},\\
&M_{52}\Rightarrow H_{62},&&M_{55}\Rightarrow H_{12},H_{31},H_{83},\\
&M_{53}\Rightarrow H_{23},&&M_{56}\Rightarrow H_{13},H_{32},H_{52},H_{61},H_{73}.
\end{aligned}
\end{equation*}
If $M_{55}$ clicks, then Charlie performs the measurement $\mathcal{M}_{5}'\equiv\{M_{51}':=P[|1\rangle_{C};|0\rangle_{c_{1}}],~M_{52}':=I-M_{51}'\}$. The outcomes corresponding to the operators $M_{51}'$ and $M_{52}'$ are $H_{83}$ and $\{H_{12},H_{31}\}$, respectively. Obviously, $\{H_{42},H_{43}\}$ and $\{H_{12},H_{31}\}$ are locally distinguishable. If $M_{56}$ clicks, we move on to the next step.

$Step~5.$ Charlie performs the measurement
\begin{equation*}
\begin{aligned}
\mathcal{M}_{6}\equiv\{M_{61}:=P[|0\rangle_{C};|0\rangle_{c_{1}}],~M_{62}:=I-M_{61}\}.
\end{aligned}
\end{equation*}
Corresponding to the operators $M_{61}$ and $M_{62}$, the subsets of postmeasurement are $\{H_{32},H_{73}\}$ and $\{H_{13},H_{52},H_{61}\}$, respectively. They are all LOCC distinguishable.

If another operator clicks in the step 1, then also a similar entanglement-assisted discrimination protocol follows.

\section{The proof of theorem 12}\label{L}
Let Alice and Bob share two EPR states $|\phi^{+}(2)\rangle_{a_{1}b_{1}}|\phi^{+}(2)\rangle_{a_{2}b_{2}}$, while Alice and Charlie share an EPR state $|\phi^{+}(2)\rangle_{a_{3}c_{1}}$.

Bob performs the measurement
\begin{equation*}
\begin{aligned}
\mathcal{M}_{1}\equiv\{&M_{11}:=P[|0\rangle_{B};|0\rangle_{b_{1}};|0\rangle_{b_{2}}]\\
                       &\quad\quad\quad~+P[|1\rangle_{B};|0\rangle_{b_{1}};|1\rangle_{b_{2}}]\\
                       &\quad\quad\quad~+P[(|2\rangle,\ldots,|d_{B}'-1\rangle)_{B};|1\rangle_{b_{1}};|0\rangle_{b_{2}}]\\
                       &\quad\quad\quad~+ P[|d_{B}'\rangle_{B};|1\rangle_{b_{1}};|1\rangle_{b_{2}}],~\\
                       &M_{12}:=P[|0\rangle_{B};|0\rangle_{b_{1}};|1\rangle_{b_{2}}]\\
                       &\quad\quad\quad~+P[|1\rangle_{B};|1\rangle_{b_{1}};|0\rangle_{b_{2}}]\\
                       &\quad\quad\quad~+P[(|2\rangle,\ldots,|d_{B}'-1\rangle)_{B};|1\rangle_{b_{1}};|1\rangle_{b_{2}}]\\
                       &\quad\quad\quad~+ P[|d_{B}'\rangle_{B};|0\rangle_{b_{1}};|0\rangle_{b_{2}}],~\\
                       &M_{13}:=P[|0\rangle_{B};|1\rangle_{b_{1}};|0\rangle_{b_{2}}]\\
                       &\quad\quad\quad~+P[|1\rangle_{B};|1\rangle_{b_{1}};|1\rangle_{b_{2}}]\\
                       &\quad\quad\quad~+P[(|2\rangle,\ldots,|d_{B}'-1\rangle)_{B};|0\rangle_{b_{1}};|0\rangle_{b_{2}}]\\
                       &\quad\quad\quad~+ P[|d_{B}'\rangle_{B};|0\rangle_{b_{1}};|1\rangle_{b_{2}}],~\\
                       &M_{14}:=I-M_{11}-M_{12}-M_{13}\}.
\end{aligned}
\end{equation*}
Charlie performs the measurement
\begin{equation*}
\begin{aligned}\mathcal{M}_{2}\equiv\{&M_{21}:=P[(|0\rangle,|1\rangle)_{C};|0\rangle_{c_{1}}]+P[(|2\rangle,\\
                                      &\quad\quad\quad\quad~~\ldots,|d_{C}'\rangle)_{C};|1\rangle_{c_{1}}],~\\
                                      &M_{22}:=I-M_{21}\}.
\end{aligned}
\end{equation*}
Similar to the proof of Theorem 11,
when $a_{1}a_{2}a_{3}$ and $b_{1}b_{2}$ are substituted for ancillary systems $a_{1}a_{2}$ and $b_{1}$ in (\ref{K1}), respectively,
    the outcomes are obtained.
It is easy to prove that these postmeasurement states are also locally distinguishable.

\section{The proof of theorem 13}\label{M}
Notice that $d_{C},d_{D}\leq d_{B}$.  The states of subsystems $C$ and $D$ are teleported  to Bob using the maximally entangled states $|\phi^{+}(d_{C})\rangle_{BC}$ and $|\phi^{+}(d_{D})\rangle_{BD}$, respectively. Their union is represented by  $\widetilde{B}$. In addition, to locally discriminate the set (\ref{24}), Alice and Bob share a maximally entangled state $|\phi^{+}(3)\rangle_{ab}$. The specific protocol is as follows.

Alice performs the measurement \begin{equation*}
\begin{aligned}
\mathcal{M}_{1}\equiv\{&M_{11}:=P[|0\rangle_{A};|0\rangle_{a}]+P[(|1\rangle,\ldots,|d_{A}'-1\rangle)_{A};\\
                       &\quad\quad\quad\quad~~|1\rangle_{a}]+P[|d_{A}'\rangle_{A};|2\rangle_{a}],~\\
                       &M_{12}:=P[|0\rangle_{A};|1\rangle_{a}]+P[(|1\rangle,\ldots,|d_{A}'-1\rangle)_{A};\\
                       &\quad\quad\quad\quad~~|2\rangle_{a}]+P[|d_{A}'\rangle_{A};|0\rangle_{a}],~\\
                       &M_{13}:=I-M_{11}-M_{12}\}.
\end{aligned}
\end{equation*}
Suppose the outcome corresponding to $M_{11}$ clicks, the resulting postmeasurement states are
\begin{equation*}
\begin{aligned}
&U_{11}\rightarrow\{|0\rangle_{A}|\xi_{i}\circ \eta_{j}\circ (0\pm d_{D}')\rangle_{\widetilde{B}}|00\rangle_{ab}\}, \\
&U_{12}\rightarrow\{|\xi_{i}\rangle_{A}|\eta_{j}\circ (0\pm d_{C}')\circ 0\rangle_{\widetilde{B}}|11\rangle_{ab}\}, \\
&U_{13}\rightarrow\{(|0\rangle_{A}|00\rangle_{ab}+|\eta_{j}^{1}\rangle_{A}|11\rangle_{ab})|(0\pm d_{B}')\circ 0\\
&\quad\quad\quad~\circ \xi_{i}\rangle_{\widetilde{B}}\}, \\
&U_{14}\rightarrow\{(|0\rangle_{A}|00\rangle_{ab}\pm |d_{A}'\rangle_{A}|22\rangle_{ab})|0\circ \xi_{i}\circ \eta_{j}\rangle_{\widetilde{B}}\}, \\
&U_{21}\rightarrow\{|\xi_{i}\rangle_{A}|d_{B}'\circ \gamma_{k}\circ \eta_{j}\rangle_{\widetilde{B}}|11\rangle_{ab}\}, \\
&U_{22}\rightarrow\{|d_{A}'\rangle_{A}|\gamma_{k}\circ \eta_{j}\circ \xi_{i}\rangle_{\widetilde{B}}|22\rangle_{ab}\}, \\
&U_{23}\rightarrow\{(|\gamma_{k}^{1}\rangle_{A}|11\rangle_{ab}+|\gamma_{k}^{2}\rangle_{A}|22\rangle_{ab})|\eta_{j}\circ \xi_{i}\circ d_{D}'\rangle_{\widetilde{B}}\}, \\
&U_{24}\rightarrow\{(|0\rangle_{A}|00\rangle_{ab}+|\eta_{j}^{1}\rangle_{A}|11\rangle_{ab})|\xi_{i}\circ d_{C}'\circ \gamma_{k}\rangle_{\widetilde{B}}\},\\
&U_{31}\rightarrow\{|d_{A}'\rangle_{A}|0\circ (0\pm d_{C}')\circ \gamma_{k}\rangle_{\widetilde{B}}|22\rangle_{ab}\}, \\
&U_{32}\rightarrow\{|0\rangle_{A}|(0\pm d_{B}')\circ \gamma_{k}\circ d_{D}'\rangle_{\widetilde{B}}|00\rangle_{ab}\}, \\
&U_{33}\rightarrow\{(|0\rangle_{A}|00\rangle_{ab}\pm |d_{A}'\rangle_{A}|22\rangle_{ab})|\gamma_{k}\circ d_{C}'\circ 0\rangle_{\widetilde{B}}\}, \\
&U_{34}\rightarrow\{(|\gamma_{k}^{1}\rangle_{A}|11\rangle_{ab}+|\gamma_{k}^{2}\rangle_{A}|22\rangle_{ab})|d_{B}'\circ 0\circ (0\\
&\quad\quad\quad~\pm d_{D}')\rangle_{\widetilde{B}}\}, \\
&U_{41}\rightarrow\{|\xi_{i}\rangle_{A}|\xi_{i}\circ 0\circ \gamma_{k}\rangle_{\widetilde{B}}|11\rangle_{ab}\}, \\
&U_{42}\rightarrow\{|\xi_{i}\rangle_{A}|0\circ \gamma_{k}\circ \xi_{i}\rangle_{\widetilde{B}}|11\rangle_{ab}\}, \\
&U_{43}\rightarrow\{|0\rangle_{A}|\gamma_{k}\circ \xi_{i}\circ \xi_{i}\rangle_{\widetilde{B}}|00\rangle_{ab}\}, \\
&U_{44}\rightarrow\{(|\gamma_{k}^{1}\rangle_{A}|11\rangle_{ab}+|\gamma_{k}^{2}\rangle_{A}|22\rangle_{ab})|\xi_{i}\circ \xi_{i}\circ 0\rangle_{\widetilde{B}}\}, \\
&U_{51}\rightarrow\{|d_{A}'\rangle_{A}|d_{B}'\circ \xi_{i}\circ (0\pm d_{D}')\rangle_{\widetilde{B}}|22\rangle_{ab}\}, \\
&U_{52}\rightarrow\{|d_{A}'\rangle_{A}|\xi_{i}\circ (0\pm d_{C}')\circ d_{D}'\rangle_{\widetilde{B}}|22\rangle_{ab}\}, \\
&U_{53}\rightarrow\{|\xi_{i}\rangle_{A}|(0\pm d_{B}')\circ d_{C}'\circ d_{D}'\rangle_{\widetilde{B}}|11\rangle_{ab}\}, \\
&U_{54}\rightarrow\{(|0\rangle_{A}|00\rangle_{ab}\pm |d_{A}'\rangle_{A}|22\rangle_{ab})|d_{B}'\circ d_{C}'\circ \xi_{i}\rangle_{\widetilde{B}}\}, \\
&U_{61}\rightarrow\{|0\rangle_{A}|0\circ d_{C}'\circ \eta_{j}\rangle_{\widetilde{B}}|00\rangle_{ab}\}, \\
&U_{62}\rightarrow\{|0\rangle_{A}|d_{B}'\circ \eta_{j}\circ 0\rangle_{\widetilde{B}}|00\rangle_{ab}\}, \\
&U_{63}\rightarrow\{|d_{A}'\rangle_{A}|\eta_{j}\circ 0\circ 0\rangle_{\widetilde{B}}|22\rangle_{ab}\}, \\
&U_{64}\rightarrow\{(|0\rangle_{A}|00\rangle_{ab}+|\eta_{j}^{1}\rangle_{A}|11\rangle_{ab})|0\circ 0\circ d_{D}'\rangle_{\widetilde{B}}\}, \\
&U_{71}\rightarrow\{|0\rangle_{A}|\xi_{i}\circ 0\circ \xi_{i}\rangle_{\widetilde{B}}|00\rangle_{ab}\}, \\
&U_{72}\rightarrow\{|\xi_{i}\rangle_{A}|0\circ \xi_{i}\circ 0\rangle_{\widetilde{B}}|11\rangle_{ab}\}, \\
&U_{81}\rightarrow\{|0\rangle_{A}|d_{B}'\circ 0\circ d_{D}'\rangle_{\widetilde{B}}|00\rangle_{ab}\}, \\
\end{aligned}
\end{equation*}
\begin{equation*}
\begin{aligned}
&U_{82}\rightarrow\{|d_{A}'\rangle_{A}|0\circ d_{C}'\circ 0\rangle_{\widetilde{B}}|22\rangle_{ab}\}, \\
&U_{91}\rightarrow\{|\xi_{i}\rangle_{A}|d_{B}'\circ \xi_{i}\circ d_{D}'\rangle_{\widetilde{B}}|11\rangle_{ab}\}, \\
&U_{92}\rightarrow\{|d_{A}'\rangle_{A}|\xi_{i}\circ d_{C}'\circ \xi_{i}\rangle_{\widetilde{B}}|22\rangle_{ab}\},
\end{aligned}
\end{equation*}
where $|\eta_{j}^{1}\rangle_{A}=\sum_{u=1}^{d_{A}-2}\omega_{d_{A}-1}^{ju}|u\rangle$,  $|\gamma_{k}^{1}\rangle_{A}=\sum_{u=0}^{d_{A}-3}$ $\omega_{d_{A}-1}^{ku}|u+1\rangle$ and $|\gamma_{k}^{2}\rangle_{A}=\omega_{d_{A}-1}^{k(d_{A}-2)}|d_{A}-1\rangle$ for $j,k\in \mathcal{Z}_{d_{A}-1}$. Evidently, they can be perfectly distinguished by LOCC. For all other cases a similar protocol follows.

\end{appendix}

\end{document}